# A training programme for early-stage researchers that focuses on developing personal science outreach portfolios

Shaeema Zaman Ahmed[1], Arthur Hjorth[2], Janet Frances Rafner[2], Carrie Ann Weidner[1], Gitte Kragh[2], Jesper Hasseriis Mohr Jensen[1], Julien Bobroff[3], Kristian Hvidtfelt Nielsen[4], Jacob Friis Sherson*[1,2]

[1] *Department of Physics and Astronomy, Aarhus University, Denmark*
[2] *Department of Management, Aarhus University, Denmark*
[3] *Laboratoire de Physique des Solides, Université Paris-Saclay, CNRS, France*
[4] *Centre for Science Studies, Aarhus University, Denmark*
*sherson@mgmt.au.dk

**Abstract**

Development of outreach skills is critical for researchers when communicating their work to non-expert audiences. However, due to the lack of formal training, researchers are typically unaware of the benefits of outreach training and often under-prioritize outreach. We present a training programme conducted with an international network of PhD students in quantum physics, which focused on developing outreach skills and an understanding of the associated professional benefits by creating an *outreach portfolio* consisting of a range of implementable outreach products. We describe our approach, assess the impact, and provide guidelines for designing similar programmes across scientific disciplines in the future.

# Introduction

Recent years have seen an increased call for outreach and science communication training by research agencies such as the National Science Foundation (National Science Foundation, 2014), the Royal Society (Royal Society, 2006), the European Commission's Research Executive Agency (European Commission, 2020) and the Vitae Researcher Development Framework (Vitae, 2011) highlighting benefits for both society and researchers. This increase can be attributed to the potential of science outreach to boost scientific literacy in society, helping citizens to counter misconceptions (McCartney et al., 2018; St Angelo, 2018), and make well-informed decisions (Brownell et al., 2013; Clarkson et al., 2014). Outreach also promotes the incorporation of science into policy-making (O'Keeffe & Bain, 2018). Furthermore, outreach initiatives conducted in schools and universities shape young students' career aspirations and introduce them to less popular and emerging scientific fields (Finan et al., 2018; Strang et al., 2005). Apart from the impact on society and young students, outreach training also benefits researchers by developing transferable skills, such as communication, presentation, leadership, and project management skills that are applicable in any career (Kuehne et al., 2014).

To address this call for training in science communication and outreach, organizations and initiatives provide workshops, short courses, and semester-long courses. These training opportunities typically target the creation of only a few outputs, developing practical skills in a particular area, or developing outreach projects. To provide some examples, the *Engage* programme (Rodgers et al., 2018) and *FameLab USA* (Scalice et al., 2019) focused on oral communication skills. The European Science Communication Network (ESConet) (Miller & Fahy, 2009) consisted of a workshop series covering theoretical and practical aspects of the social science and media aspect of science communication. Furthermore, the *Physics Reimagined* team (Bobroff & Bouquet, 2016) conducted semester-long courses where students designed an outreach project. Lastly, organizations such as the American Association for the Advancement of Science (AAAS) (Basken, 2009) and the Alan Alda Center for Communicating Science (Bass 2016) provide toolkits and resources to researchers through seminars and credit-based courses.

Although these independent training opportunities exist, researchers rarely have access to formal outreach training as it is seldom a part of institutional curricula in Science, Technology, Engineering, and Mathematics (STEM) (Clarkson et al., 2014; Mellor, 2013; O'Keeffe & Bain, 2018; Ponzio et al., 2018; Trautmann & Krasny, 2006). Studies reported that lack of time, insufficient financial resources, and institutional bureaucracy hinders the development of training opportunities on the institutional side (Davies, 2013; Noble et al., 2016; Royal Society, 2006; Wortman-Wunder & Wefes, 2020). These challenges arise due to the fact that the perceived importance of science communication and outreach is relatively low and thus overlooked both by institutions and researchers (Ponzio et al., 2018; St Angelo, 2018). Moreover, little academic merit is gained from engaging in outreach (Neeley et al, 2014; Davies, 2013), and this hinders the motivation of researchers to do outreach during their career (Triezenberg et al., 2020).

Therefore, although there is a desire for researchers to play an important role in science communication and outreach, they may not be motivated or equipped to engage in outreach alongside their research career. This is problematic because communication with a general audience requires training to incorporate deliberate practise and careful attention to many aspects, e.g. language and context, to help researchers distil and simplify complex scientific information to make it understandable for a general audience (Brownell et al., 2013).

In order to address the lack of knowledge about the professional benefits of outreach for researchers and the deficit of formal training in developing and applying a wide variety of outreach tools and resources, we present an outreach training programme that helped PhD students gain outreach skills and increase their understanding of career benefits through an outreach portfolio-based approach. In the context of our training, we define an outreach portfolio as a collection of outreach products, personal outreach motivation, philosophy, and experiences, analogous to teaching and design portfolios. In comparison with existing independent training interventions that focus only on one or two outputs and are typically not integrated along with the trainees' study or research timeline, the novelty of our programme lies in the following: (a) pluralistic approach of creating a range of implementable outreach products during a PhD-training timeline and (b) building an outreach portfolio for real-world use.

The objectives of our training programme were the following:

- Facilitate students' learning about and skills relating to outreach through designing and creating a suite of outreach products for the public, high-school students and undergraduates.
- Provide learning and reflection experiences so that students can experience which kinds of outreach products best suit their personal needs, skills, tastes, research and outreach settings.
- Increase students' awareness and understanding of the transferability of outreach skills to other domains of their careers.

In this paper, we first present an overview of related work, focusing on other outreach training programmes for students or early-stage researchers. We identify typical learning goals and design features of these other programmes to argue for the novelty of our programme design. We then describe our training programme, and the activities and outreach products that participants created during the programme. Finally, we present an assessment of the outcomes of the training programme addressing two research questions:

RQ 1: How and what did students learn and enjoy from participating in the workshops?
RQ 2: What were students' perceptions and learning from creating their outreach products?

This study was conducted through surveys completed by students and senior researchers at the end of the training programme. Based on our experiences gained in this training programme, we provide a list of guidelines for designing and implementing similar future programmes to train and support early-stage researchers' communication skills.

## Outreach Training Initiatives

To position our training programme within the context of other outreach training initiatives and establish the novelty of our programme, we outline and categorize existing training opportunities reported in previous studies based on the type and aim of the training. Specifically, we look at the form and duration of such efforts, and the learning goals emphasized. We acknowledge that there most likely exist other training opportunities that have not been reported on.

As far as form and duration, training opportunities typically consist of the following types:

- Seminars (Bishop et al., 2014; Clarkson et al., 2014; Gau et al., 2020; Gianaros, 2006; Trautmann & Krasny, 2006; Triezenberg et al., 2020)
- Intensive short courses (Fogg-Rogers et al., 2020; Rodgers et al., 2018)
- Semester-long courses (Bobroff & Bouquet, 2016; Clarkson et al., 2018; Crone et al., 2011; Heath et al., 2014; Marbach-Ad & Marr, 2018; Ponzio et al., 2018; St Angelo, 2018)
- Workshops ranging from one hour to five days in duration (Andrade Oliveira et al., 2019; Holliman & Warren, 2017; McCartney et al., 2018; Miller & Fahy, 2009; O'Keeffe & Bain, 2018; Scalice et al., 2019; Strang et al., 2005; Trautmann & Krasny, 2006; Webb et al., 2012; Wortman-Wunder & Wefes, 2020)

As far as learning goals in other existing programmes, students practice written and oral communication skills, as well as managerial and pedagogical competencies, which are described below.

*Written communication*
Workshops on written communication train participants in writing blog posts or popular science articles either about one's own research or general topics connected to the researchers' field (Bishop et al., 2014; Heath et al., 2014; Rodgers et al., 2018; O'Keeffe & Brian, 2018, Wang et al., 2018).

*Oral communication*
Training in oral communication focuses on elevator-style pitches and presentations, where, for example, graduate students participate in a science communication event and deliver three-minute oral presentations (Scalice et al., 2019; Rodgers et al., 2018). Additionally, some programmes focus on storytelling and culminate in the delivery of 20-60 minute oral presentations for the general public (Clark et al., 2016; Clarkson et al., 2018; Gau et al., 2020).

*Designing and implementing project-based outreach products*
Project-based courses enable trainees to develop an outreach product, such as creating an experimental setup for live demonstrations or animated movies (Bobroff & Bouquet, 2016), interactive booths (Crone et al., 2011), and exhibits (Webb et al. 2012). Some programmes also train students or provide support to researchers in the implementation of outreach activities in collaboration with local organizations (Heath et al., 2014; St Angelo, 2018; Triezenberg et al., 2020; Webb et al., 2012).

*Educational outreach*

Educational outreach initiatives train graduate students to conduct activities with middle-school students such as teaching them about peer-review with practical exercises (Trautmann & Krasny, 2006), and organizing guided lab tours for middle- and high-school students (Clark et al., 2016). Additionally, a few programmes also provide training to graduate students and researchers to conduct outreach in science classrooms, such as annotator training aimed at making research papers accessible to undergraduate students (McCartney et al., 2018).

*Our training context*

To summarize, other training programmes typically focus only on one or two learning experiences and outputs that can be used in specific settings. Additionally, very few programmes exist that are spread out over a longer timeline (e.g. a 2-3 year PhD or graduate study timeline) and are integrated with the trainees' study or research timeline. Shorter timeline programmes are relatively informal and voluntary. These typically introduce the basics and benefit only a subset of motivated participants whereas longer timeline programmes are coursework-based formal training that provide more skills and confidence (Brownell et al., 2013).

Furthermore, outreach training interventions help researchers to gain transferable skills as categorized by Kuehne et al. (2014) that can be beneficial in their career (Table 1), but researchers may not necessarily be aware of this due to a lack of emphasis on the importance of professional benefits of outreach by peers, supervisors, and even universities (Davies, 2013; Kuehne et al., 2014, Heath et al., 2014; Miller & Fahy, 2009, Ponzio et al. 2018).

**Table 1:** Transferable skills of outreach training.

| Author, Year | Key Learning Goals: |
|---|---|
| **Improved communication skills** | |
| Marbach-Ad & Marr, 2018; Noble et al., 2016; Webb et al., 2012; Trautmann & Krasny, 2006 | <ul><li>Improved ability to explain one's research and its importance in oral and written tasks</li><li>Better communication with people outside their field</li><li>Improved skills in writing grants and science magazine articles</li></ul> |

| **Improve presentation and engagement skills** | |
| --- | --- |
| Clarkson et al., 2018;<br>Heath et al., 2014;<br>Marbach-Ad & Marr, 2018;<br>Scalice et al., 2019;<br>Ponzio et al., 2018;<br>Rodgers et al., 2018;<br>Triezenberg et al., 2020 | ● Improved storytelling<br>● Better presentation skills<br>● Using graphics, thematic, and structural elements to visualize complex data in posters and presentations<br>● Better connection with the audience<br>● Improved engagement skills<br>● Better understanding of engagement methodologies |
| **Improved teaching skills** | |
| Fogg-Rogers et al., 2015;<br>Strang et al., 2005;<br>Trautmann & Krasny, 2006 | ● Increased knowledge about pedagogy and teaching<br>● Higher awareness of the challenges of teaching<br>● Gaining methods for improving teaching skills<br>● Knowledge about factors influencing the ways students learn<br>● Increased content knowledge outside own field |
| **Improved career skills** | |
| Bishop et al., 2014;<br>Crone et al., 2011;<br>Gau et al., 2020;<br>Holliman & Warren, 2017;<br>McCartney et al., 2018;<br>O'Keeffe & Bain, 2018;<br>Scalice et al., 2019;<br>Wortman-Wunder & Wefes, 2020 | ● Increased self-efficacy and confidence<br>● Bolstering résumé's for early-stage researchers<br>● Advantages in job interviews and gaining better industry positions<br>● Improved networking skills |
| **Improved leadership and management skills** | |
| Trautmann & Krasny, 2006;<br>Triezenberg et al., 2020 | ● Improved organization and time management skills<br>● Increased skills in facilitative leadership to lead students and peers better<br>● Fostering partnerships for advancing science |

Therefore, in our training programme, the approach was to (a) provide a learning experience through the hands-on creation of outreach products that can be used in a variety of settings and tailored according to a diversity of participants' needs and goals, (b) emphasize the professional benefits of outreach skills, and (c) implement the programme over a longer time frame in order to integrate it with the students' PhD timeline. This provided opportunities and ample time for practice, sustainable progress, and ultimately the creation of a range of usable outreach products. Our approach built upon previous studies on best practices on the emphasis of feedback, quality of outputs, and infusing training into the curriculum (Brownell et al., 2013; Silva & Bultitude, 2009). In the following section, we will describe our training programme in more detail.

## The QuSCo Outreach Training Programme

The training programme was created and run by the ScienceAtHome team (https://www.scienceathome.org/) based at Aarhus University for the Intensive Training Network **Qu**antum-enhanced **S**ensing via Quantum **Co**ntrol (QuSCo), funded by the European Union's Horizon 2020 Research Innovation Programme (CORDIS, 2020). The QuSCo network is a research network group consisting of 15 PhD students and 23 senior researchers working on different domains in quantum physics research, out of which one student (S.Z.A) and project leader (J.S.) from the network are part of the ScienceAtHome team and authors of this manuscript.

*Goals*

The training programme was mandatory for all the PhD students as it was a QuSCo work package with the following goals: (a) development of outreach tools (simulations/graphical illustrations/popular science article/games) and (b) development of outreach activities for different settings (lab tours/workshops/science fairs). In order to achieve these goals, we designed a series of workshops with outreach product creation activities spread out over the students' PhD timeline to provide students with training aimed at developing their outreach skills and creating their personal outreach portfolio.

*Programme design*

The training programme ran over the span of 23 months from March 2019 to January 2021, and consisted of 11 sessions: seven 2-hour workshops, two talks, and two online events where students presented their work (Figure 1). Due to different start and end months of the PhD students within the QuSCo timeline (November, 2017-2021), the training programme started after all the PhD students were enrolled and ended before any student completed their PhD. All the sessions were conducted online except three, which were conducted as physical workshops since they were part of the conferences organized by the QuSCo network.

At the end of the workshops, students created their personal outreach portfolios consisting of six outreach products created during the course of the programme: (a) Blog, (b) Infographic, (c) Poster, (d) 15-second "My Research" video, (e) Simulation, and (f) Inquiry-Based Learning (IBL) worksheet.

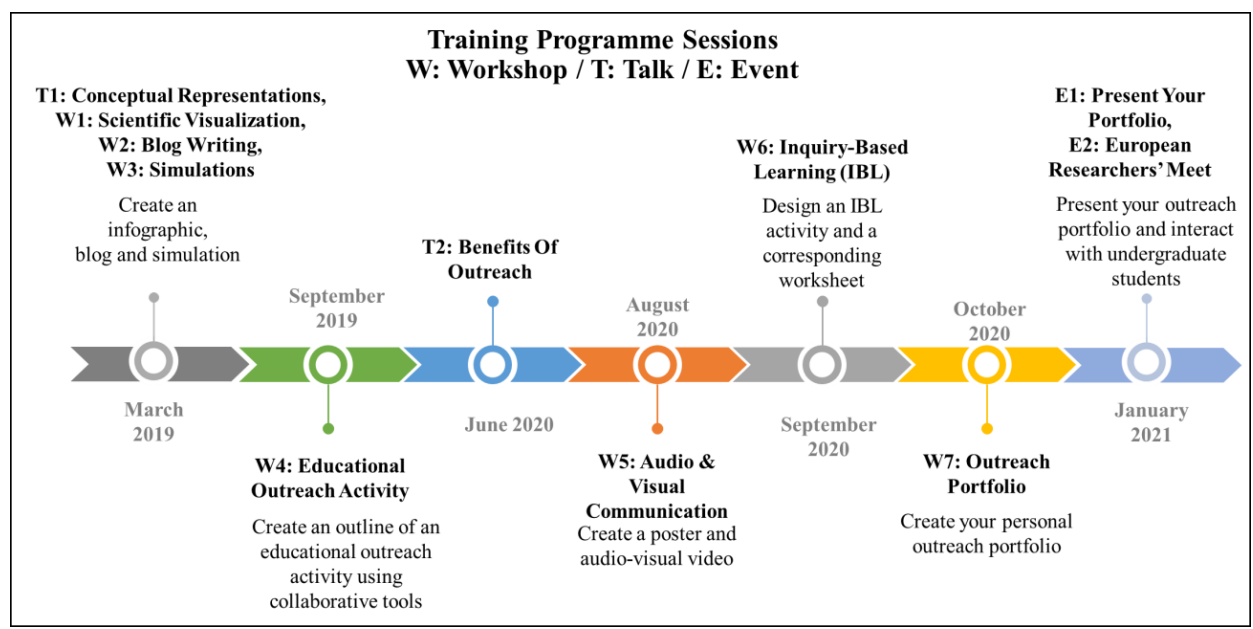

**Figure 1**: Timeline of sessions of the QuSCo training programme. The training programme consisted of 11 sessions: seven 2-hour workshops (W), two talks (T), and two online events (E) that took place between March 2019 and January 2021.

Each session had a theme, learning goals and a corresponding task for students (Table 2). The talks served as an introduction to the themes covered in the workshops. For instance, the talk on *Conceptual Representations* was a basic introduction to how efficient conceptual communication can stimulate learning of complex concepts, which underlined the motivation for the blog writing and infographic workshops that followed. Similarly, the talk on *Benefits Of Outreach* gave students a perspective on the importance of outreach and alignment with the training programme.

**Table 2**: Summary of tasks and goals for students in the training programme.

| Month Year | Session: W: Workshop / T: Talk / E: Event | Student tasks | Goals for students |
|---|---|---|---|
| March 2019 | T1: Conceptual Representations<br><br>W1: Scientific Visualization<br><br>W2: Blog Writing (physical workshop) | ● Create an infographic<br>● Write a blog<br>● Create a 1-minute simulation | Practice:<br>● how to write for a non-academic audience.<br>● giving insight into your research using drawings and images. |
| | W3: Simulations (physical workshop) | | ● highlighting the core of your research through a simulation tool. |
| September 2019 | W4: Educational Outreach Activity (physical workshop) | ● Create an outline of an educational outreach activity using collaborative tools | ● Gain experience with adapting existing tools in designing outreach activities that communicate relevant aspects of their research |
| June 2020 | T2: Benefits Of Outreach | ● Fine-tuning previous work | ● Provide a sense of the societal, professional and personal benefits acquired through outreach, and highlight that practice and feedback will be an integral component of the workshops |
| August 2020 | W5: Audio and Visual Communication | ● Design a poster<br>● Create a 15-second "My Research" video where students talk about their research in a catchy and fun format. | ● Practice creating an output that conveys a message in a short amount of time through audio and visual communication |
| September 2020 | W6: Inquiry-Based Learning (IBL) | ● Design an IBL activity and a corresponding worksheet | ● Provide an IB learning experience in order to |

| | | | implement a scientific, inquiry-driven educational outreach activity |
|---|---|---|---|
| October 2020 | W7: Outreach Portfolio | ● Create a slideshow of their portfolio<br>● Upload and update their portfolio products on the QuSCo website | ● Emphasize the importance of having an outreach portfolio<br>● Participate in a brainstorming session to reflect on the students' motivation to do outreach and their philosophy |
| January 2021 | E1: *Present Your Portfolio*<br><br>E2: *European Researchers' Meet* | ● Students present their outreach portfolio to the panel of senior researchers<br>● Students interact with undergraduate students by giving a 5-minute talk | ● Participate in real settings to showcase their portfolio and outreach products |

Each workshop consisted of the following training phases: (a) Homework Tasks, (b) Theory, (c) Brainstorm, (d) After-Session Tasks, (f) Peer Feedback, and (g) Expert Feedback, as shown in Figure 2. Homework Tasks refer to any preparatory work needed *before* a workshop, e.g. watching a video, which was done before the session. Phases (b) and (c) were done in-session, after which the students completed an activity or product in the *After-Session Task*s and received peer and expert feedback over a period of 2-3 weeks. After the completion of all phases, the products were uploaded on QuSCo's website (https://qusco-itn.eu/outreach-3/) and/or YouTube channel, as applicable.

The feedback process was iterative and consisted of an initial round of peer feedback, followed by a round of expert feedback from the research team consisting of learning scientists, science communicators, education researchers, and quantum physicists. Furthermore, during peer feedback, pointers were given to the students by the instructors to outline what students should consider while giving feedback. For all the text-based tasks such as writing a blog or creating an IBL worksheet, feedback was typically given through comments and suggestions in *track changes* mode such that students could see the feedback concretely and iteratively improve their tasks. For the graphical or visual tasks, preliminary feedback was given during the workshop, followed by a second round of feedback as a list of suggestions or comments communicated through email.

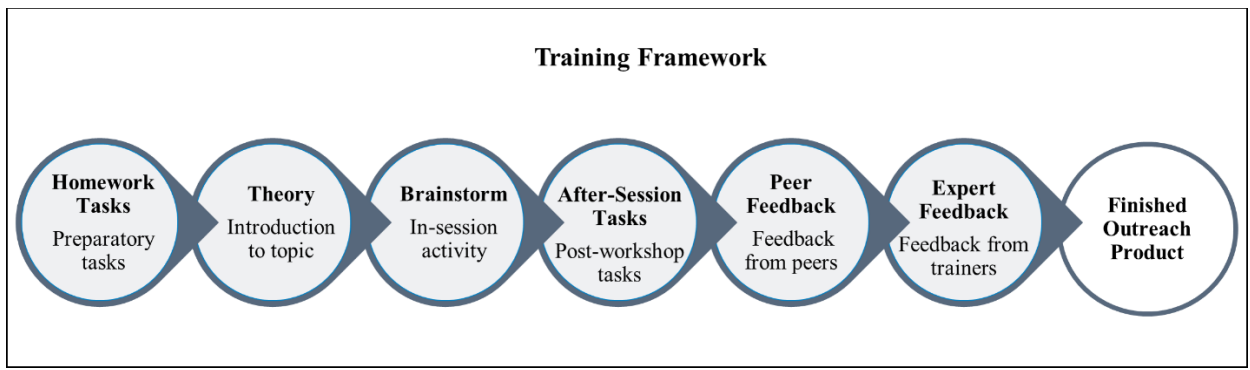

**Figure 2**: Our training framework consisted of six training phases deployed in each workshop that culminated in a finished outreach product.

We emphasize that the focus of the programme was on facilitating students' *creation* of the products, while *implementing and testing* their products was left to students' own interests and opportunities in their outreach settings. However, to provide opportunities to prepare students for future real-world outreach, the programme culminated in two events, one where students presented their outreach portfolio to the board of senior researchers in the network at the *Present Your Portfolio* event and another where students gave 5-minute talks to undergraduate physics students at the *European Researchers' Meet* organized by the ScienceAtHome team. Students were encouraged to use their products in the mentioned outreach scenario and some of them indeed used their portfolio products such as the infographic, poster, and the simulation at the *European Researchers' Meet* event, as well as in other outreach events such as the Munich Conference on Quantum Science and Technology 2020, where one of the QuSCo students won an award for the best poster (Nymann, 2020) Students took initiative to test and spread their products on their own. Many posted the 15-second "My Research" video on Twitter and one students' video was selected in the *#MyJobinResearch* video challenge on Twitter (Research Executive Agency, 2020).

*Workshop Descriptions*
The QuSCo training programme structured the seven two-hour workshops in the following order: (i) *Scientific Visualization*, (ii) *Blog Writing*, (ii) *Simulations*, (iv) *Educational Outreach*, (v) *Audio and Visual Communication*, (vi) *Inquiry-Based Learning*, and (vii) *Outreach Portfolio*. The workshops covered a wide variety of outreach aspects, such as written and visual communication, designing an outreach activity, creating a learning track, and emphasizing the importance of outreach in one's career. Workshops were designed to provide a combination of theoretical background and an activity for students to perform, observe or experience. The aim of the activities was to help students reflect or further work on concrete tasks. In the following, we present a brief overview of the primary goal and content of each workshop. A more detailed account of each workshop can be found in Appendix A (under Supplementary Material).

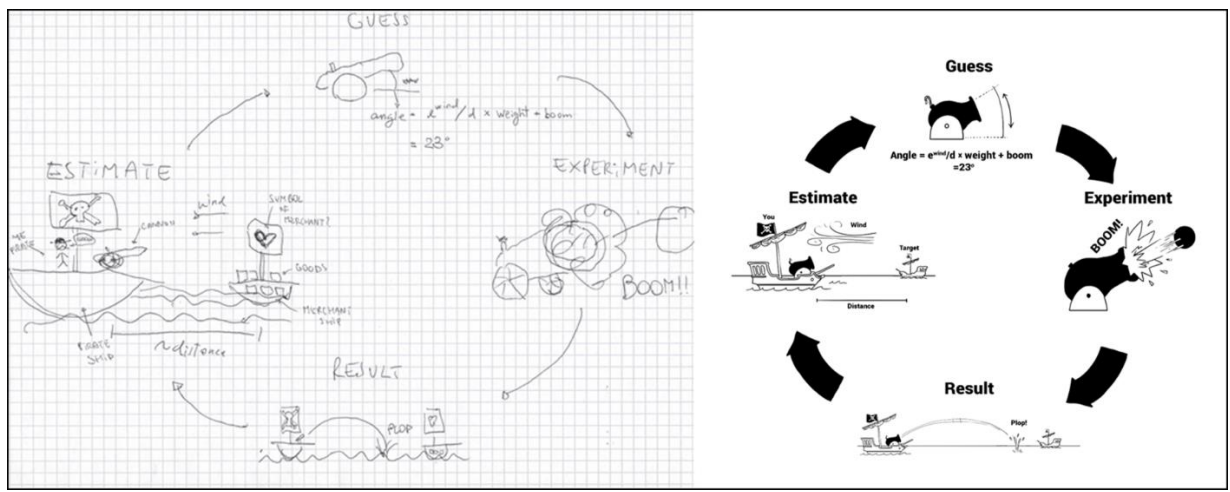

**Figure 3**: Initial draft of the infographic (left) made by Federico Roy from the QuSCo network and the final product (right) after a professional artist rendered it digitally. The infographic depicts the iterative nature of an optimal control technique used in quantum physics research.

The *Scientific Visualizations* workshop focused on visual communication that included a presentation that drew on Gestalt Principles (Koffka, 1935) of visual design to teach how people typically perceive information contained in graphics. This was followed by a sketching session where students sketched an infographic on paper that was then sent to a professional artist to produce a digital version (Figure 3). The *Blog Writing* workshop first provided strategies for effective written communication in blogs targeted at high-school students, such as using analogies and the active voice (Baram-Tsabari & Lewenstein, 2012), followed by a writing session where students outlined the content, overarching headlines, and take-away message of their blog.

As simulations are an integral component in teaching and learning quantum physics (McKagan et al., 2008, Kohnle et al., 2012, Weidner et al., 2020) and since QuSCo is a network of quantum physics researchers, we dedicated two workshops to creating and using simulations for educational outreach activities. The *Simulations* workshop aimed at supporting students in creating bite-sized videos that showed a dynamic visualization of a core quantum mechanical phenomenon used in their research that could be used for dissemination and teaching activities with undergraduate physics students. The tools used for creating the simulation were *Quantum Composer* (Zaman Ahmed et al., 2020) and *SpinDrops (*https://spindrops.org/) created by two research groups within the QuSCo network. Furthermore, another workshop using *Quantum Composer* together with *SciNote*, a collaborative scientific argumentation tool (Rafner et al., 2021) *was* conducted to inspire students and help them develop an outline of their own activity in the *Educational Outreach* workshop. This workshop consisted of students first experiencing an activity around a quantum physics challenge, which provided students with ideas to create their own activity using collaborative educational and research tools. The final output was the outline of a specific use-case for using collaborative learning tools for outreach, e.g. addressing programming challenges using IBM's Quantum Experience (https://quantum-computing.ibm.com/) together with *SciNote*.

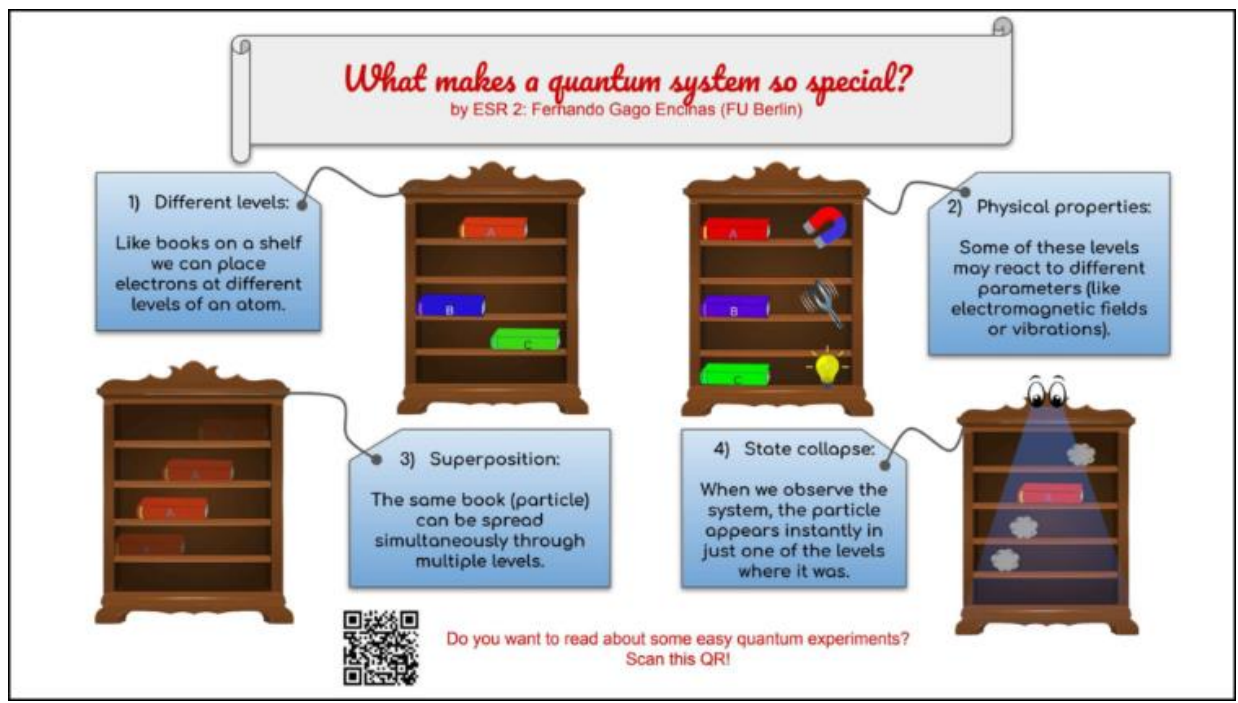

**Figure 4**: Screenshot of Fernando Gago-Encinas's (QuSCo PhD student) poster describing the properties of a quantum system through analogies.

In the *Audio and Visual Communication* workshop, students were introduced to communicating concepts through a poster and a video that conveys their research or any concept in a very short time, which can be used in science exhibitions or social media. The goal of the workshop was to provide the foundation for making a poster and a video of themselves talking about their research. The video task was inspired by the European Commission's *#MyJobinResearch* video challenge on Twitter (Research Executive Agency, 2020), where students had to speak about their research in 15 seconds. For inspiration, we showed graphics and videos from the *Physics Reimagined* outreach team (http://hebergement.universite-paris-saclay.fr/supraconductivite/). During the workshop, we discussed the guidelines on structuring their poster inspired by *Compound Interest (*https://www.compoundchem.com/*)*. We also provided tips on including graphical elements such as pictures, their own infographics, and a QR code to embed information (such as their simulation or blog) for further details. Additionally, in a practice task, students were divided into groups and each student explained their research in 15 seconds to the others in their group. The final output was a poster (Figure 4) and a video created by the students that became part of the outreach portfolio.

In order to train students to teach a concept from their research to high-school and undergraduate students, the *Inquiry-Based Learning* (IBL) workshop focused on introducing students to a learning framework based on a scientific inquiry-driven process and creating a learning track that could be used for outreach in classrooms and informal learning settings. The workshop began with an activity to experience the IBL process where students discussed and provided hypotheses to answer a driving question (unrelated to the students' research fields), followed by a brainstorming activity where students found data online, posed questions and provided arguments in support of their hypotheses. After this activity, we showed a concrete example of an IBL worksheet built around a quantum mechanical concept using *Quantum Compose*r, one of the simulation tools used in previous workshops. The activity and the example from the workshop laid the foundation for students to create their own IBL activity and a corresponding worksheet. The intended aim of creating the worksheet was to help students consider learners' prior knowledge, both by addressing potential sources of misconception and by thinking concretely about how to unfold learning in outreach settings conducted in a classroom.

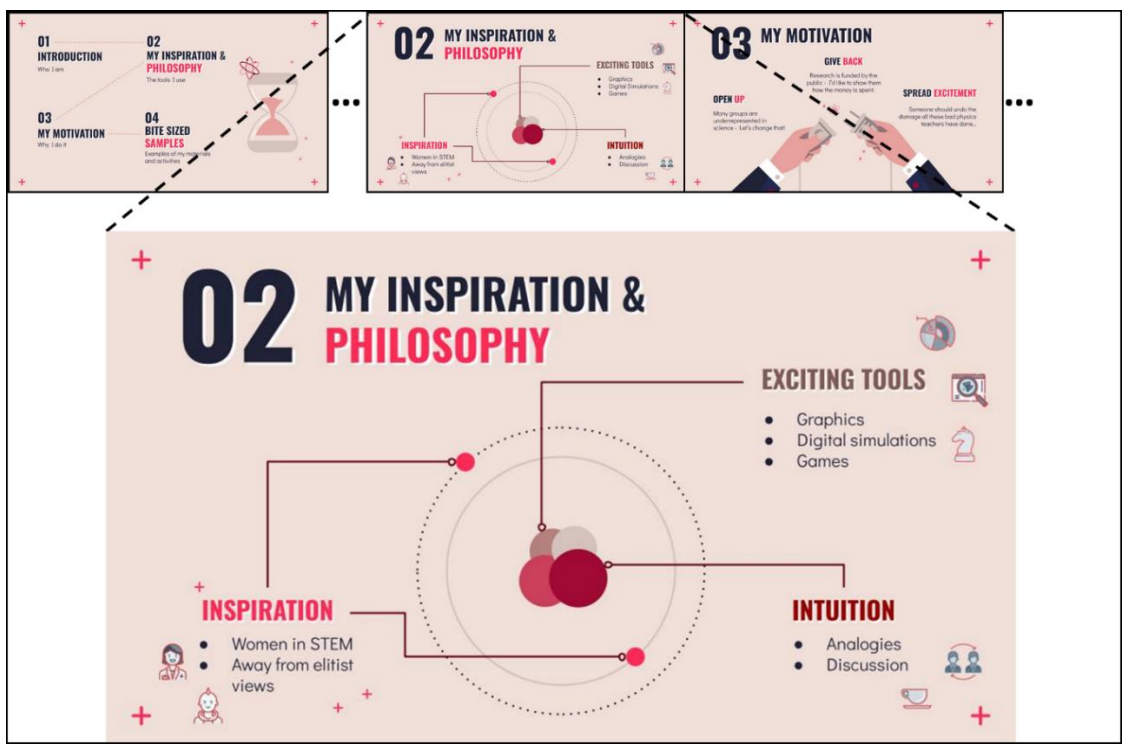

**Figure 5:** Collection of screenshots of Phila Rembold's (QuSCo PhD student) outreach portfolio slideshow. The top three images depict some of the various components covered in the portfolio. The bottom image shows the enlarged view of one of the components describing personal inspiration, philosophy, and tools for outreach.

The last workshop, *Outreach Portfolio,* laid the foundation for curating an outreach portfolio describing the students' outreach skills and competencies; this portfolio could be useful for future job applications. In the workshop, we discussed and crowdsourced students' thoughts on outreach philosophies and the motivational factors that implicitly helped students to reflect upon their personal attitudes and preferences while doing outreach. The intention was to provide students with a sense of accomplishment by displaying the outreach products they created over the course of the programme and emphasize the use-cases of their portfolio for job applications, similar to teaching and design portfolios. After the workshop, the students created their personal outreach portfolio as a set of slides (Figure 5) and a dedicated personal page on the QuSCo website. Subsequently, students presented their portfolios in a simulated 'job interview' at the *Present Your Portfolio* event setting where each student had five minutes to convince a simulated 'hiring committee' consisting of the board of QuSCo senior researchers and two external reviewers about their outreach skills and experiences. Following this simulated interview, the students received feedback and comments from the committee.

## Methods: Data Collection and Analysis

In order to assess the outcomes of the training programme, we gathered data from students and senior researchers of the QuSCo network through surveys. The student and senior researcher surveys are attached in appendices B and C, respectively (under Supplementary Material). Surveys consisted of a mix of four-point Likert-scale questions (1; *Strongly disagree* - 4; *Strongly agree*), multiple-choice questions, and open-ended questions. We chose an even-numbered set of options for all Likert-scale questions in order to avoid middle-of-the-road responses. Thus, all answers were either in some degree of agreement or some degree of disagreement with the statement in the prompt.

The purpose of the Likert-scale questions was to generate numerical data for assessing participants' overall thoughts on the programme and the learning activities. The purpose of the open-ended questions was to get elaborations on students' responses to the Likert-scale questions. The survey contained a mix of two kinds of questions and had a 200-character minimum limit on all open-ended questions. The student survey was sent to 14 students, out of which 12 completed the survey. The senior researcher survey was sent to 23 researchers, out of which 16 completed it.

Surveys were sent to participants a month after the final workshop, and student participants were given a week to fill them in. This deadline ensured that students filled in the surveys before the *European Researchers' Meet*, as the survey contained questions related to the student's apprehension or excitement regarding the event. Senior researchers were given a deadline that was a week after the *Present Your Portfolio* event in order to respond after observing their students doing outreach at the portfolio presentation event.

Surveys were sent out using SurveyXact (https://www.surveyxact.com/). We exported student responses to Microsoft Excel for recording codes and creating plots. Three members of the team did the qualitative analysis, taking the following approach: two members (S.Z.A and A.H) first met and articulated focal points for the coding of responses to each question based on our research interest in students' responses. For instance, for the survey question, "What was most difficult about communicating your research effectively with non-physicists?", the focus was established to be "what specific things did students report that were difficult?". The two researchers then coded all responses individually. They then met to discuss their codes, and how they had applied them to individual responses with the aim of converging on a coding scheme and a fully coded dataset. In this conversation, some codes were removed and some were combined. Based on these codes, we wrote a codebook (Table 3) and asked a third member (paid research assistant) to code all responses. Finally, we calculated Cohen's Kappa (Cohen, 1960; Warrens, 2015) between the two coded datasets to establish inter-rater reliability. As our codes were not mutually exclusive (i.e. one student's response to a question could contain several codes), we calculated a Kappa for each code individually. Thereafter, we calculated the average and overall Kappa across all codes in our dataset, which were found to be 0.75 and 0.71 respectively.

**Table 3:** Codebook and code frequencies for the open-ended questions in the student survey. The codebook was divided into sets of codes (A-G), each corresponding to a topic covered in the student survey.

A. **Experience:** Responses that describe an experience that students had as a result of a training step.

| *Codes* | *Statement focus* | *Responses coded* |
| --- | --- | --- |
| Reflection | Reflection of any thoughts or methodologies | 1 |
| Togetherness | Feeling of togetherness in the programme | 2 |
| Improvement | Improvement of a product or work | 7 |
| Learning from feedback | Learning gained from peer or expert feedback | 10 |
| Self-confidence/ Encouragement | Gaining self-confidence and/or encouragement | 2 |
| Peer perspectives | Hearing peer perspectives | 2 |
| Inspiration | Being inspired | 3 |

B. **Unexpected learning:** Responses that describe something that students express they were unaware or surprised to know or learn about.

| *Codes* | *Statement focus* | *Responses coded* |
| --- | --- | --- |
| Interactive medium | Use of games and interactive research tools in outreach | 1 |
| Time/Effort | Understanding how much time and/or effort is needed to prepare for outreach | 1 |
| Teaching techniques | Learning inquiry-based learning and other teaching techniques | 4 |
| Transferable benefits | Understanding that outreach skills can be beneficial or help gain transferable skills in one's career | 5 |

C. **Request for more skills/topics:** Responses that specify topics and/or skills that students would have liked more time for and/or learned more about.

| *Codes* *(S:Skill / T:Topic)* | *Statement focus* | *Responses coded* |
|---|---|---|
| S: Interviewing | Interviewing skills (e.g. being a science journalist) | 1 |
| T: Outreach online | How to conduct outreach or present in online scenarios | 1 |
| S: Public speaking | Skills in public speaking in talks or presentations | 2 |
| T: Teaching approaches | Different teaching approaches or methods | 1 |
| T: Theory, impact, and history | Learning about the history and theoretical principles of outreach, and the impact of outreach on society | 2 |
| S: Organize an outreach event | Practical skills on how to organize an outreach event | 2 |
| S: Professional production | Use of professional tools for making graphics and producing audio and visual content | 5 |
| S: Audience interaction | Interacting with and tackling different audiences, e.g. high-school students, industry partners, and the public | 4 |
| S: Writing | Skills in different writing styles and approaches | 1 |
| Spending more time | Asking to spend more time on making any particular outreach product | 4 |
| T: Pros and Cons of product types | Assessing the advantages and disadvantages of different product types while interacting with different target audiences | 1 |

D. **Proud of a product**: Responses that describe *what* made students feel proud of a product they made (excluding responses related to participating in the workshops or the programme).

| *Codes* | *Statement focus* | *Responses coded* |
| --- | --- | --- |
| Captures thoughts | Product captures or displays a student's imagination, idea or research concept | 3 |
| Real-world use | Product that can be used in a real-world setting with the public or students | 4 |
| Feel-good | Product that is fun, interesting, poetic, or visually appealing | 4 |
| Good feedback | Received good and encouraging feedback | 1 |
| Overcoming challenges | Proud of overcoming challenges, e.g. simplifying content or concepts while making the product | 2 |
| Time/Effort | Spending time and/or effort in doing something | 3 |
| Simplification | Product that was successful in breaking down concepts into simpler terms, analogies or illustrations | 6 |

E. **Product revision**: Responses that specify *what* changes or revisions students would make to a product (excluding things related to what they did not like about the product).

| *Codes* | *Statement focus* | *Responses coded* |
| --- | --- | --- |
| Time | Would like to just spend more time on something | 1 |
| Look/Sound better | Change something to make it look or sound better and nicer | 3 |
| Redo | Redo something after having learned something new | 2 |
| Add content | Adding content or being more descriptive in a product | 2 |
| Understandable | Making something more understandable | 2 |
| Alignment | Revise something to align with other portfolio product(s) | 1 |
| Pick another topic | Change the topic covered in the product | 1 |

F. **More training for product revision:** Responses that describe the need for more time, training, or support to help students revise the products that they felt needed more work.

| *Codes* | *Statement focus* | *Responses coded* |
| --- | --- | --- |
| No training | Students state that they do not need any more training | 7 |
| Practical experience | Asking for more practical implementation or experience | 4 |
| Examples | Asking to see more examples | 1 |
| Professional production | Asking for training in using professional tools for video, graphic, or audio production | 3 |
| Time | Time is an issue/hindrance | 4 |
| Simplification | Asking for more training on simplifying concepts | 1 |

G. **Difficulties:** Responses that describe what students found difficult about communicating their research with non-physicists using their products.

| *Codes* | *Statement focus* | *Responses coded* |
|---|---|---|
| Time constraints | An activity or product that by design has to be performed in a short time | 3 |
| Audience considerations | Assessing the audience's prior knowledge and what the audience would like to know or prefer to learn | 4 |
| Engagement | Building curiosity or engaging the audience in some way | 2 |
| Simplification | Simplifying complex concepts to make it more understandable for the audience | 4 |

# Results

In this section, we present the results from the student and senior researcher surveys in two subsections, covering each of the surveys respectively.

## Student Survey

We present the data organized under two overarching research questions:
    RQ 1: How and what did students learn and enjoy from participating in the workshops?
    RQ 2: What were students' perceptions and learning from creating their outreach products?

We answer these two questions by presenting findings from our surveys. We report results based on numerical data from responses to Likert-scale questions with elaborations from open-ended questions where present. Regarding the elaborations, student quotes and the corresponding response number and code(s) are indicated.

**RQ1: How and what did students learn and enjoy from participating in the workshops?**
The first part of the question regarding student learning was assessed through three questions on the survey, and the last regarding student enjoyment through two questions.

1. **Which training phases were most useful?**

We wanted to understand which particular training phases students found most useful. Therefore, we asked students to select any two phases (amongst six) they found most useful from the training phases described earlier in the QuSCo Training Programme section.

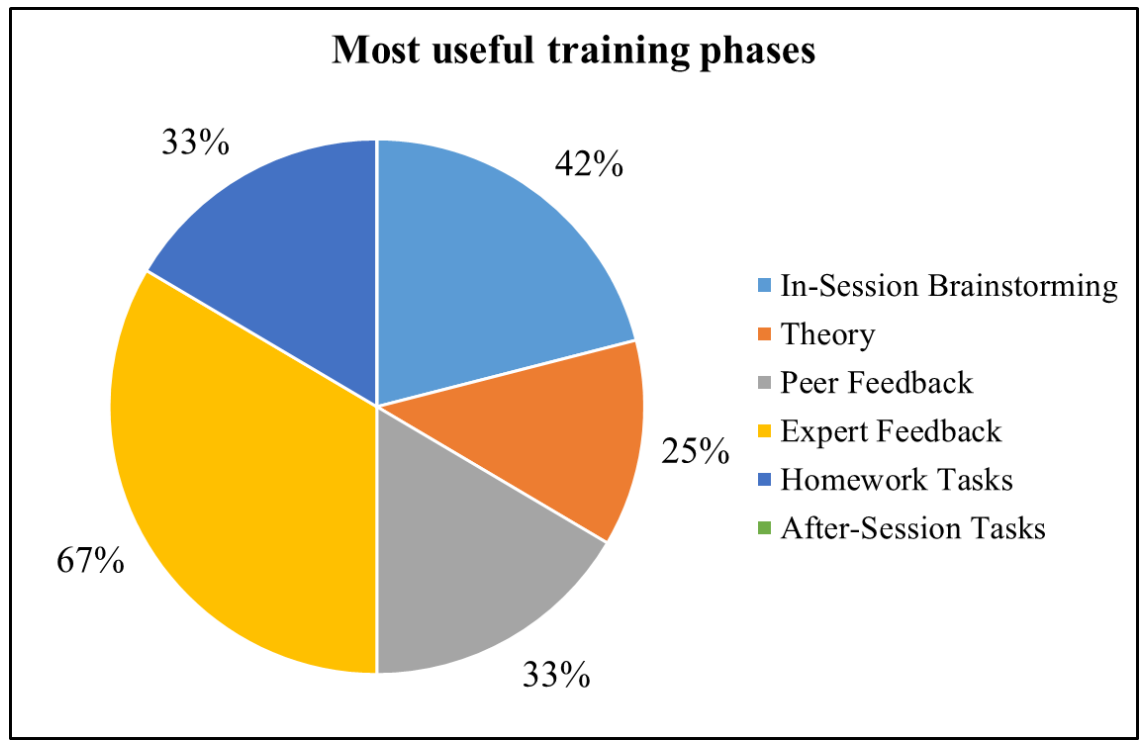

**Figure 6**: The most useful training phases, as self-reported by students. As students chose two phases, the total is 200%. We note the large variety in students' selection of training phases.

We identified two primary takeaways from the student responses to this multiple-choice question (Figure 6): (i) students found expert feedback to be very useful, and (ii) all phases of the training were found to be useful, except *After-Session Tasks*.

It is notable that no students chose *After-Session Tasks*. We believe that the students may have confused this element with *Homework tasks*, or believed it to be the same element, and therefore did not select it. We think that the element was not defined clearly by us (S.Z.A and A.H) at the time of writing the survey, which resulted in this confusion.

This multiple-choice question was followed by an open-ended question that asked students to describe an *experience with one of the training phases that they enjoyed*. Here, 10 out of 12 students emphasized that they particularly enjoyed learning from feedback, and 7 out of 12 emphasized that they enjoyed improving their products. Students stated that expert feedback in particular helped them in a multitude of ways, e.g. by correcting mistakes, identifying potential sources of misunderstanding, and gaining new perspectives while creating their outreach products (e.g. blog, poster, IBL worksheet). The following quote exemplifies one such response.

> *"As I didn't have any previous experience with outreach activities, I found it difficult to design outreach products, such as the blog and the learning activity. Feedback from experts was extremely valuable, both in adjusting my language (in the case of the blog) to be more accessible, and in improving my planned questions, in the learning activities."* [R4; Learning from feedback, Improvement]

**2. In what ways did the programme meet students' expectations and provide learning?**

To understand students' prior expectations and their alignment with the training programme, we asked students whether there were things in the programme that surprised them or things that they did not expect to learn about. In our analysis, two things stood out: (i) about half of students had not previously considered outreach skills to be generally applicable to their careers (5 out of 12) and (ii) a number of students explicitly mentioned the IBL activity had been an interesting learning experience (4 out of 12). The following quotes exemplifies the two aspects respectively:

> *"At the beginning of my outreach training here in the QuSCo network, I personally was not feeling confident in talking about my research fields, topics and interests. Moreover, I was even less confident in talking about those issues to non-scientific audience*[s]*, to non-experts. By now, after these workshop series, I feel like I could explain quantum mechanics to my grandmother.*
>     *At the beginning of my outreach training here in the QuSCo network, I thought that outreach was not very important for my career in academia or outside academia. I thought that research is all about theorems and experiments, and that there would be no time for anything alse* [else] (sic)*. By now, after these workshop series, I have realised that the ability of speaking about your research, advertise your scientific results, which also means conveying excitement in science to other people, is as much important as your results."*
> [R5; Transferable benefits]

> *"I liked very much the inquired-based learning seminar and the work we have done to create our inquired-based worksheet. I didn't expect something like this and I really appreciated it [...........], I found very useful and intersting* [interesting] (sic) *doing the inquired based learning wrksheet* [worksheet] (sic). *And I think it is an outreach activity that I will use in the future."* [R6; Teaching techniques]

3. **Which skills or topics would students like to spend more time on or learn more about?**

To inform us about improvements in future training programmes, we asked students if there were any skills or topics that they would have liked to spend more time on, or skills or topics that were not covered in the workshop but that they would have liked to learn about.

Two particular topics stood out. Five students out of 12 stated that they would have liked to learn about practical aspects of video, sound, and graphical design through professional tools. This was not covered in our workshops, and participants just worked within whichever software they were already familiar with or could easily access. Four out of 12 also asked for more training on how to interact and deal with the target audience. Some of these gave specific examples like high-school students, or potential industry partners.

While not exactly an answer to the question, four students reported that they just would have liked to have more time. This is noteworthy because it means that even an extensive programme like ours left some students wanting to spend more time on outreach.

We present a few examples from the mentioned findings:

> *"I quite liked the 15s video challenge. It would have been great to go over some basics like lighting, picking a background, regulating the microphone etc or get access to software that can be used to produce subtitles for hearing impared* [impaired] (sic) *people or insert graphics. I would have also liked to learn how to use a more professional graphics building tool than powerpoint.*
> *What I would have liked to learn is which difficulties one might run into when interacting with Highschool* (sic) *students and how to deal with them. I would have also appreciated a workshop on how to check whether the level of explanation is appropriate* […….]. *I would have also liked to actually practice speaking i.e. make a full length outreach presentation and hold it."* [R8; S: Public speaking, S: Professional production, S: Audience interaction]

> *"I think some tutorials for making designs would have been really nice (on Inkscape to keep it easy)!....."* [R10; S: Professional production]

> *"I'm very satisfied with the teachings in the workshop. Maybe focusing on a specific activity, for example the enquiry-based* (sic) *learning, and actually organizing an event were* [where] (sic) *we all made our activity at a 1 day event."* [R4; Spending more time]

4. **Did students enjoy participating in the workshop series?**

Students were asked whether they enjoyed the workshop series and if they would recommend this experience. Results from the student surveys show that all students enjoyed the workshop (Figure 7a). Additionally, all students except one reported that they would recommend the workshop series to other PhD students (Figure 7b).

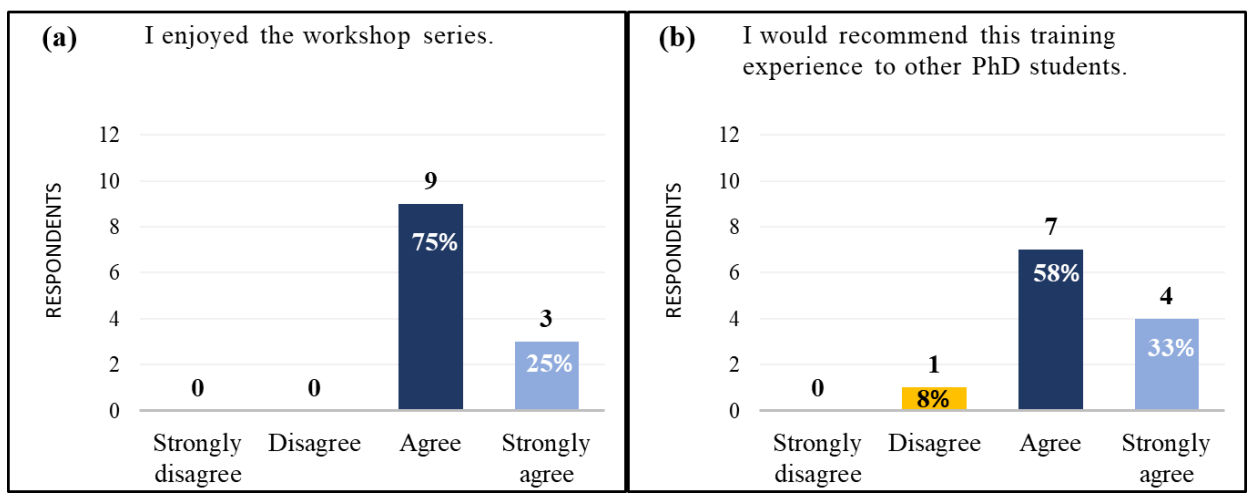

**Figure 7**: Results from survey questions focusing on whether students enjoyed the training programme. Most students enjoyed the training programme and would recommend the programme to other PhD students.

**RQ2: What were students' perceptions and learning from creating their outreach products?**

As a core part of our intervention focused on participants' creation of outreach products, we also asked about students' perception of their products and what they learned from making them.

1. **Which portfolio product were students proud of?**

    We asked students which portfolio product they were particularly proud of, and which product they felt needed more work, and what they felt it needed. There was a wide variety of students' choices in selecting the outreach product they were most proud of (Figure 8). All products were chosen by at least one student.

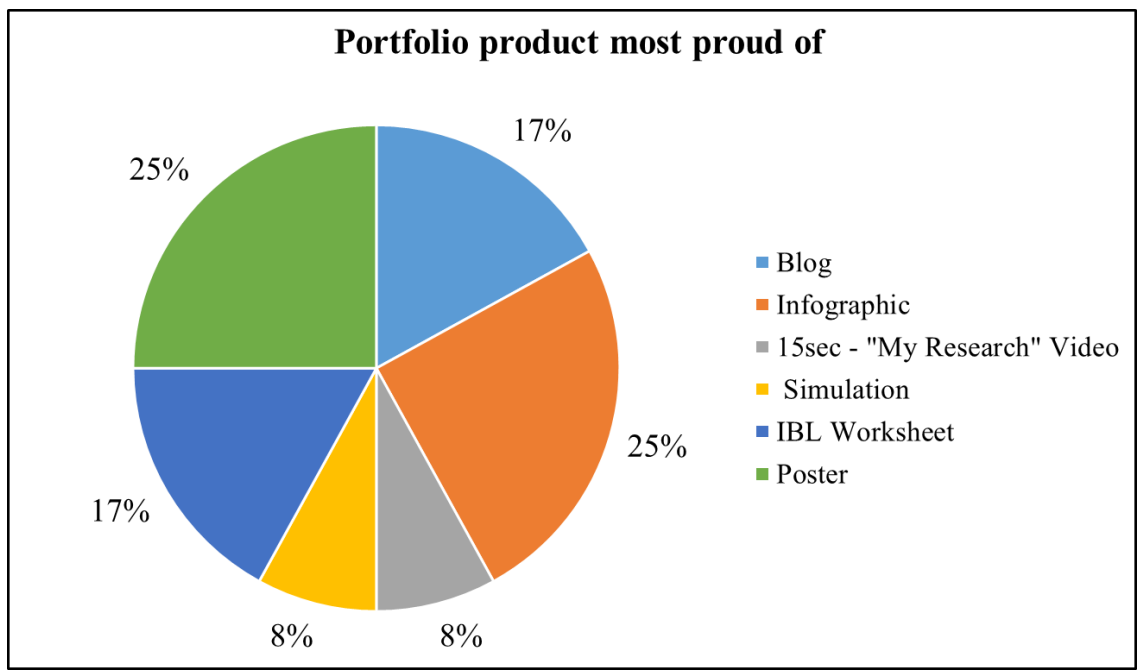

**Figure 8**: Results of the student responses corresponding to the survey question, "Which portfolio product are you most proud of?". We note a wide variety in students' selection of products.

To understand students' responses better, we also asked them to describe *what* makes them proud in an open-ended follow-up question. Six out of 12 students specifically mentioned that they were proud of having explained complex research in simple terms. Students' responses indicated that four students were proud of their outreach product because they thought it was a fun and an interesting product, and also four students stated that designing for real-world use cases was motivating. To elaborate on these results, we present some responses:

> *"I think I was able to achieve two goals* [with the blog]*:*
> *- simplify my topic to be made understandable to the general public*
> *- make the blog funny and interesting enough to hopefully lead to further questions."* [R4; Feel-good, Simplification]
>
> *"For me, the Blog is the outreach product that I am most proud of because it was a challenge for me to explain my research work in simpler terms […], and I think I succeeded in that. And I am pretty sure that this is also going to help me in the* *future. I also liked the fact that it went through rigorous feedbacks* (sic).*"* [R5; Simplification, Overcoming challenges]

*"I used materials which I had considered in my head for a long time. It was nice to see these things coming together. This* [poster] *is also the only product which people actually get to see (as I have put it on my office door)."*
[R8; Real-world use, Captures thoughts]

2. **What did students feel about needing more training to revise their products?**

We wanted to know whether students felt that they needed more training in any particular skills. To make this as concrete as possible, we asked students which of their products they felt needed more work, then asked them *what* about the product needed more work, and finally whether they felt they had the skills required to do this, or if they needed more training. Figure 9 shows that at least one student indicated each product, with the exception of the *Blog*. In a follow-up question, we asked what they would like to change or revise in the product but the responses varied widely, and no clear patterns emerged from students' responses to this question.

Finally, as mentioned we asked them what kind of additional training they would need. Four out of 12 students mentioned that they need more practical experience with using their products. Additionally, three students reported needing training in professional tools for video, audio, or image production. Seven students reported that they do not need further training, and four students stated that they simply needed more time to revise or practice. This does not exactly answer the open-ended question but since a considerable number of students reported this, we think it is worthwhile to present it as it points to another important theme of time and outreach.

The following quotes exemplify the mentioned findings:

> *"I think I have the skills* [to work on the poster]*, but it recquires* (sic) *lots of time. For sure, if I had more practice with some design program it would have been easier. But I think that it is essentialy* (sic) *the lack of time the problem."* [R7; No training, Practical experience, Time, Professional production]
>
> *"I would need more video editing skills* [for the simulation]. *It would also be nice to talk about ways to make videos easier to watch on the phone and* *talk to someone from the traget* [target] (sic) *group what they would like to have changed."* [R8; Professional production]

*"Yes, I think I need more training, I need to work with a real audience and implement my IBL worksheet. In theory it can work but in practice it might go wrong. I hope I will have this opportunity in the nearest future."* [R12; Practical experience]

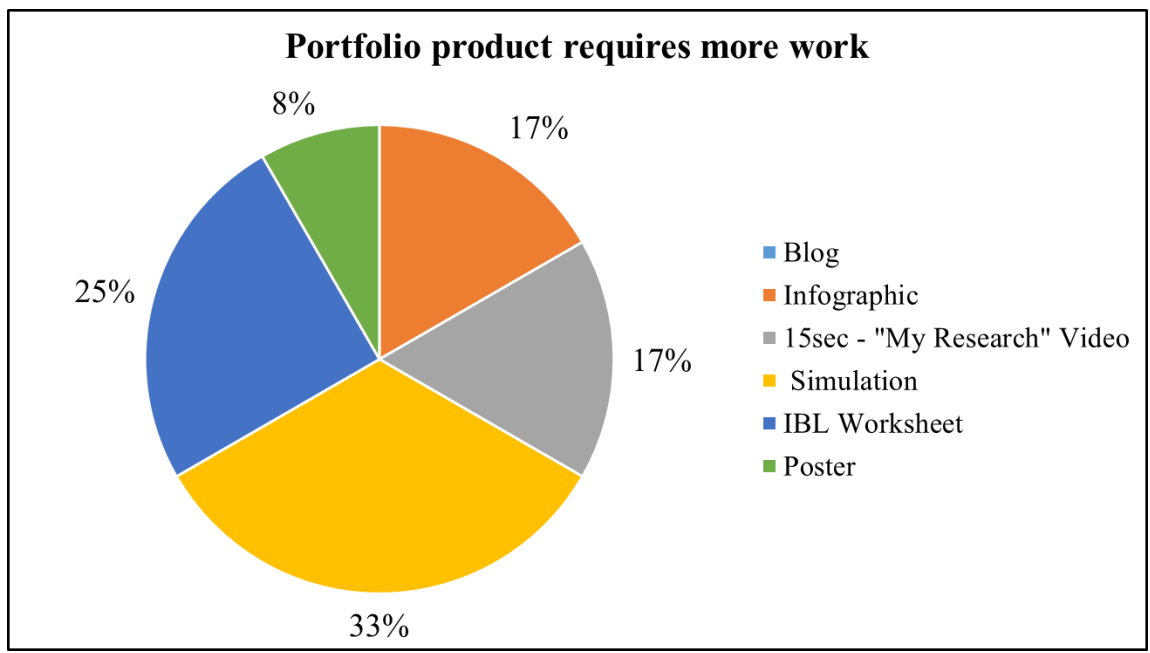

**Figure 9**: Results of the student responses corresponding to the survey question, "Which portfolio product needs more work?". At least one student indicated each product, except the *Blog*.

3. **Which portfolio product was most difficult for students to design so that it effectively communicated with non-physicists?**

As we wanted to know what students found difficult to design for outreach purposes, we asked them which portfolio product was the most difficult to design for effective communication with non-physicists. There was a wide variety of students' choices in the responses (Figure 10). All products were chosen by at least one student, except the *Infographic*.

a. **What was difficult about designing the product?**

Four out of 12 students reported that they found it most difficult to simplify complex topics or concepts. Further, four out of 12 students reported that understanding an audience, either their preferences for science communication or their prior knowledge, was the most difficult aspect of designing a portfolio product. Finally, three students (out of four who reported the 15-second video as the difficult product to design) mentioned that the constraint of 15 seconds made it the most difficult.

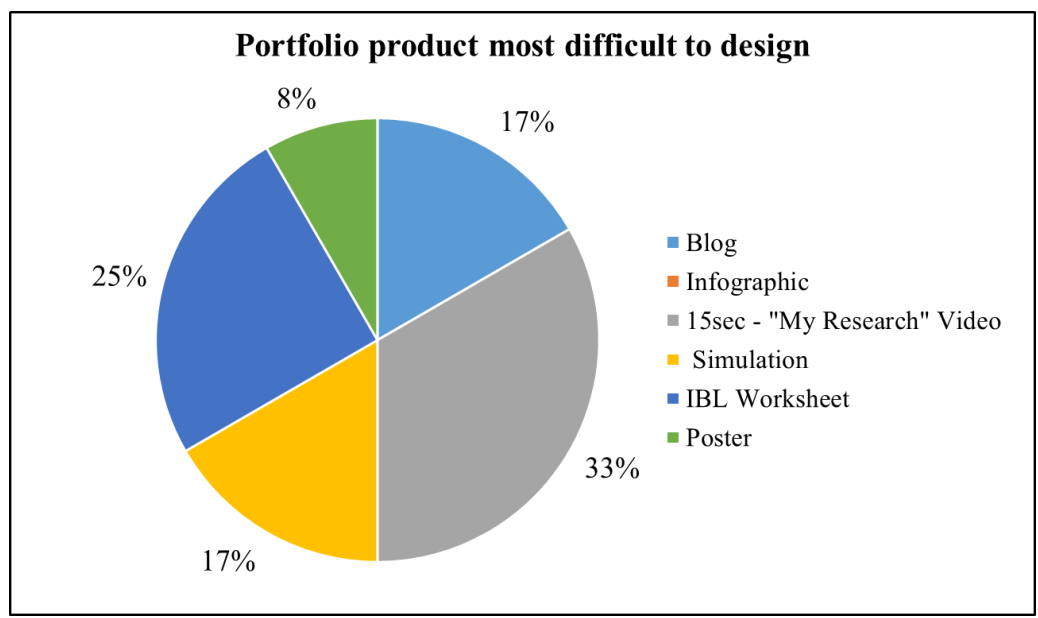

**Figure 10**. Results of the student responses corresponding to the survey question, "Which part of your portfolio was most difficult to design so that it effectively communicates with non-physicists?". At least one student reported each product, except the *Infographic*.

We present three quotes exemplifying the findings:

> *"How to introduce the experimental set-up and explain how a two levels system works* [in the simulation]. *I do not know how much it is necessary to say in order to be able to well explain the Physics* (sic) *behind a two levels system whitout* (sic) *being repetitive"*
> [R7; Simplification]

> *"It's hard to foresee what non-physicists consider important and fun to learn about* [in the 15sec- "My Research" video]." [R2; Time constraints, Audience considerations]

> *"It is really difficult to explain anything considerable to a non-expert in 15 seconds* [in the 15sec- "My Research" video]*, even by using props or any other medium."* [R1; Time constraints]

**4. How did students benefit from the variety in outreach activity product types?**

As one overarching aim of the training programme was to expose students to a variety of approaches to creating outreach products and help students realize which kinds of outreach products personally work for them, the survey consisted of questions concerning students' perception and level of confidence with respect to creating and implementing their own outreach products. All but one student agreed that they have a better sense of which types of outreach products and formats they personally want to use in their own outreach settings in the future (Figure 11a).

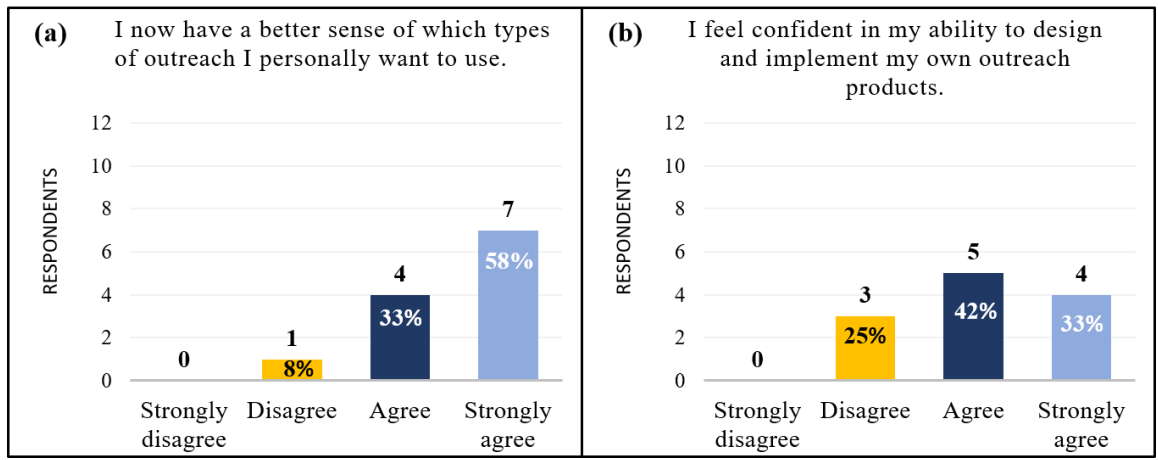

**Figure 11**: Results from survey questions concerning students' perception of implementing outreach products. Most students had a better sense of which types of outreach suit their needs and some students were confident in implementing the products.

However, when students were asked whether they felt confident in their ability to design and implement their own outreach products, the results showed more disagreement among the students with three students indicating they did not feel confident (Figure 11b). Furthermore, all students except one, who strongly agreed to question (a), either agreed (N=2) or strongly agreed (N=4) in question (b). This indicates that most students who strongly agreed to having a better sense of which types of outreach products they would personally use were also confident to design and implement the products.

**5. What are students' thoughts on using their outreach portfolios in their future career?**

As our intention in the programme was also to help students understand the benefits of outreach efforts in a variety of settings that may arise in their career, we asked them to answer questions concerning their thoughts on the use-cases and benefits of an outreach portfolio. Most students agreed that outreach skills and products are generally useful in their career (Figure 12) to perform better academically (Figure 12a) and in job interviews or communicating with other physicists (Figure 12b).

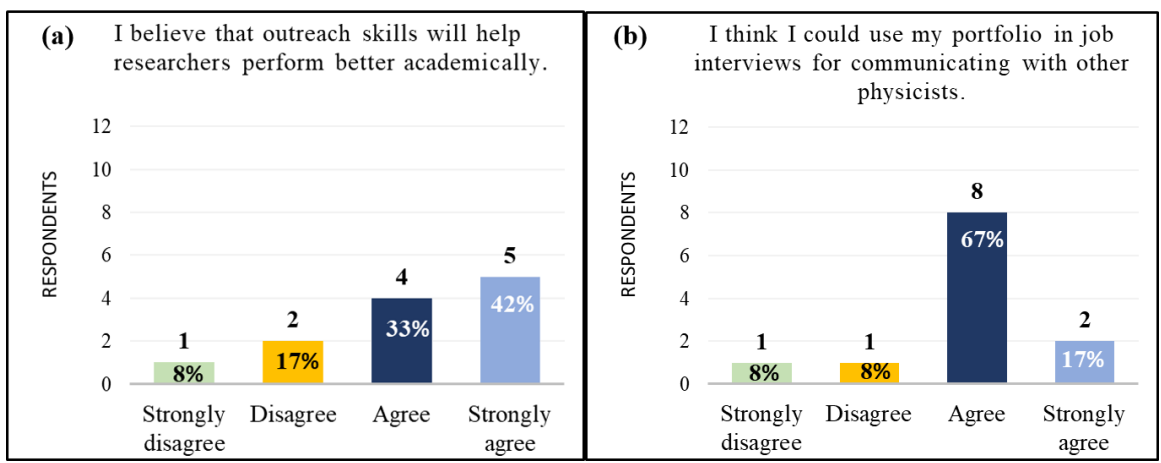

**Figure 12**: Results from survey questions focusing on whether students believed that outreach skills and their portfolio could be used more generally in their careers, outside of outreach work. Most students agreed that outreach skills and products are generally useful in their career.

## Senior Researcher Survey

In addition to asking students to answer a survey, we also asked the board of senior researchers in the QuSCo network to answer a separate short survey to explore their views on outreach training. While the objective of this survey does not directly address our research questions, we wanted to assess the senior researchers' perceptions on the training programme, portfolio products and institutional support for outreach to explore setbacks and potential areas of intervention in the future.

The majority (N=14) out of 16 senior researchers agreed or strongly agreed that the creation of the outreach portfolio in the training programme was a good learning experience for students that allowed them to improve their outreach skills (Figure 13a). Slightly fewer senior researchers (N=12) agreed or strongly agreed that it should be spread out over students' PhD timeline (Figure 13b), and ten senior researchers agreed or strongly agreed that it should be mandatory for students to be part of outreach workshops (Figure 13c). Lastly, the majority of senior researchers (N=11) agreed or strongly agreed that researchers were well supported to take part in outreach in their organization, though five senior researchers disagreed with this statement.

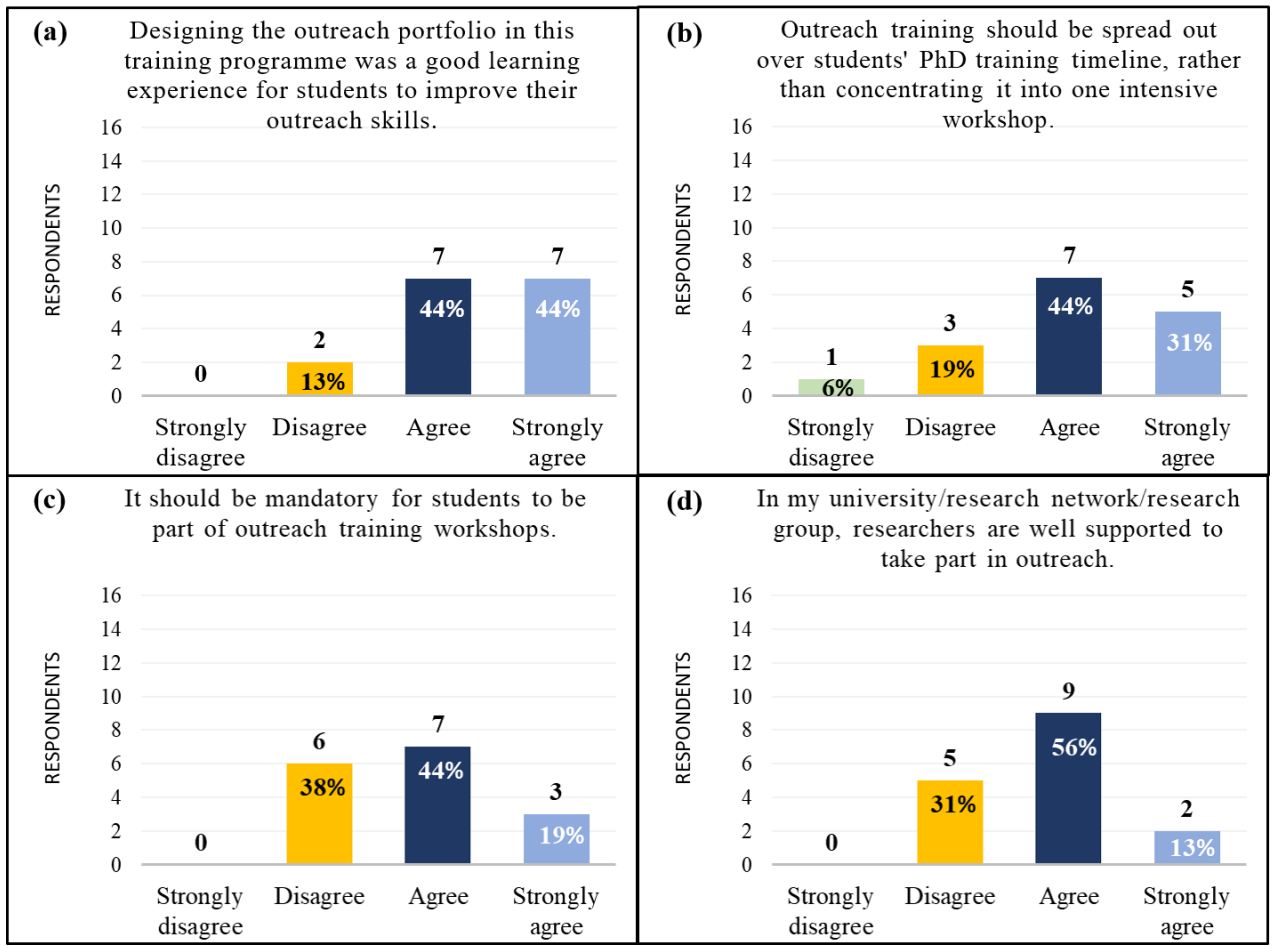

**Figure 13**: Results from the senior researcher survey, focusing on their assessment of the QuSCo training programme, outreach training in general, and of the institutional support for outreach at their respective institutions.

## Discussion

Based on the results from student and senior researcher surveys, we here discuss the achievements and shortcomings of our training programme and the creation of a personal outreach portfolio by the students. We discuss what outcomes students gained as a result of the training programme, views of senior researchers on outreach training, and the need for additional interventions in the future to meet student needs as well as to make outreach inclusive as part of academic life.

*Training programme*
Our training programme included a diversity of training phases (e.g. feedback, brainstorming, etc.) and our results show that students considered most of them useful. This highlights how important it is to include a variety of training phases in programmes and ensure sufficient time and resource allocation for each phase from both students and senior researchers. Expert feedback was particularly highly valued as it helped students learn considerably more and improve their products. This underlines the importance of committing expert resources to give in-depth feedback to students during training programmes, as also found in the *Communicating Science* course (Ponzio et al., 2018).

As one of our objectives was to make students aware of the transferability of outreach skills to other domains in their career, another key finding was that students reported that the programme made them aware of the potential benefits that outreach training or skills can bring to their careers. Moreover, some students were unaware of this before the training programme, which suggests an issue relating to the importance of outreach in academia in general, as also mentioned in previous studies (Ponzio et al., 2018; St Angelo, 2018; Triezenberg et al., 2020). Additionally, our interpretation is that when training programmes expose students to many different outreach products, it helps students to select which products to use based on their personal preferences and use case settings.

Regarding the results from the senior researcher surveys, our interpretation is that the senior researchers were generally satisfied with the training programme, though some disagreed that outreach training workshops should be mandatory for students. While we did not ask for elaboration on answers, this might suggest that some senior researchers believe that outreach training should be a purely personal choice by students or academics and that outreach is not a core part of an academic career. This could pose a challenge for students or researchers involved in outreach, as it might be perceived as outside the domain of regular academic work, and therefore done during "free time" and not recognized as valuable. This further confirms previous study findings that suggested that institutions and supervisors under prioritize outreach efforts, which affect the involvement of early-stage researchers in outreach activities (Davies, 2013; Kuehne et al., 2014, Heath et al., 2014; Miller & Fahy, 2009, Ponzio et al. 2018). However, the fact that the majority of senior researchers recognized the value of outreach is encouraging, and gives us hope that outreach may be more highly valued in the future. Moreover, most students reported that they did not need further training but expressed the need to find time alongside their research for outreach. To us, this emphasizes that a training programme like ours cannot stand on its own, but needs institutional backing and support from supervisors to ensure that students are given enough time to learn and practice outreach. This will also benefit PhD students who may pursue a career outside academia where outreach or transferable skills learned during outreach training and practice could be valuable to their careers.

*Outreach portfolio*

As our pluralistic approach to creating outreach products is one of the primary innovations of our programme, a key finding was that there was a wide variety in students' choices of the outreach product they were most proud of. Hence, we believe that training programmes ought to include a wide range of different products in order to serve a wide student population. Students reported that making products for real-world use cases was motivating. We think this points to another important design decision in our training programme that students created products that were intended for real use. We believe that creating for real-world use cases is more fruitful than asking trainees to learn through designing products that only serve as classroom exercises.

The outreach portfolio developed at the end of the training programme provided most students with a sense of accomplishment and a useful way to display their personal reflections, outreach products, skills, and experience, such that students could use the portfolio in job interviews or applications. Furthermore, most students agreed that they had a better sense of which types of outreach products they personally wanted to use. This indicates that the training programme achieved the intended goal of exposing students to a wide variety of outreach tools and resources in order to provide them with a sense of which outreach products they feel comfortable with and enjoy using in their own contexts.

To our knowledge, the creation of an outreach portfolio has not been employed in other outreach training programmes. Based on our results, we believe that this approach is a beneficial way for students to create a range of usable outreach products and a tangible way to highlight outreach skills, efforts and experiences that can be useful for early-stage researchers in their career.

*Additional interventions*

The surveys also revealed shortcomings of our programme, and the results will inform the future direction for this programme as well as guidelines for new programmes.

Firstly, the programme did not extensively cover the practical aspects of outreach and our assessment showed that students would have liked more practical implementation guidance for them to revise their products according to their chosen audience. Our interpretation is that students realized what kind of outreach tools and formats they would personally like to use, but students may not necessarily feel confident in their own ability to implement their products in an outreach activity.

Secondly, students also reported that assessing the audience's prior knowledge and tailoring the breakdown of complex concepts accordingly was one of the difficulties they encountered while creating their products. Novices in outreach often find it difficult to limit scope and focus on simple topics, because they find it more difficult to assess what is necessary (Brownell et al., 2013). In future versions of our training programme, we will emphasize this aspect of the design process even more, and scaffold the learning for students as they choose the scope of their products.

Lastly, students reported that they needed training in using professional tools for audio-visual and graphical production, e.g. software to edit or design graphics or videos used by professionals such as artists or designers. The exposure to professional production tools could be important for creating outreach products, though it may be difficult to afford the licences or find the relevant instructors for this purpose within a research group or university. One of the solutions to this challenge could be to introduce template-based tools such as *Canva* (https://www.canva.com/) that students can customize to create posters or flyers. To give an example from the programme, we introduced *Slidesgo* (https://slidesgo.com/)*,* a website that consists of customizable PowerPoint templates that were used by the students during the *Present Your Portfolio* event. However, this was introduced towards the end at the last workshop on outreach portfolio, and therefore students had not explored it extensively. Another solution could be to conduct tutorials of free or open-source graphics softwares with relatively shallow learning curves such as *Inkscape* (https://inkscape.org/)*,* as suggested by one of the students, or *Photopea* (https://www.photopea.com/) to overcome this challenge. Lastly, we suggest that PhD programmes or research groups could ensure that students have access to such training in their university either through collaborations, similar to the support we provided with a professional artist to digitally render the students' infographics. Another concrete example is the *Physics Reimagined* team that uses the approach of creating outreach material through collaborations with designers and graphic artists, but this requires dedicated funding to pay the professionals for their time.

Overall, the results suggest that the training programme has been a beneficial experience for the students. However, due to the low number of participants in this study (N=12 students, N=16 senior researchers), the inferences drawn may not be sufficient to make any statistically significant claims about enhanced outreach skills. Similarly, our lack of data from before the programme began makes it difficult for us to state with high accuracy the degree to which students' responses are a consequence of participating. Students often wrote about their knowledge or attitudes before starting the programme in their open-ended questions, but these responses can be subject to recall bias.

*Guidelines for designing similar programmes in STEM*
Taking into account the achievements and shortcomings of our training programme and the guidelines for best practice in STEM communications training (Silva & Bultitude, 2009), we propose that other research groups or universities could adapt the training programme. While this programme was implemented for students studying quantum physics, the key methods, findings, and conclusions are domain-agnostic since our programme consisted of only one session focused on quantum physics (the workshop on quantum simulation tools). Therefore, by making alterations to our programme and including discipline specific themes or tools relevant to the trainees' field(s), our training framework is adaptable to other scientific disciplines using the following guidelines:

- Assess prior knowledge of trainees to know the baseline to cater to participants' needs as well as for formative and summative evaluation.
- Inform participants and relevant project leaders (e.g. supervisors, senior researchers, project coordinators of the students) of the intended goals, time and effort investment, and what awaits them in the training. Relating this with our training programme, in hindsight, the workshop on the outreach portfolio should have been conducted at the beginning rather than at the end of the programme so that both students and senior researchers would be prepared for it and value it to be productive.
- Support and prioritization from supervisors or institutions is crucial.
- Training should be a collection of phases including theory, practical exercises, feedback and homework assignments to be useful for as wide a selection of students as possible.
- Include brainstorming activities in training sessions. To give an example from the training programme, the brainstorming session on outreach philosophies and motivation generated a rich list of thoughts that helped students to reflect.
- Wherever possible, include an example activity for the participants to experience and reflect upon, followed by a demonstration conducted by the participants themselves.
- Include opportunities for improvement through a feedback process, especially by including a panel of experts to provide feedback and make sure to provide ample time for the participants to update their outputs.
- Ensure contact with real 'outreach' audiences, e.g. undergraduate students in your own university. This will provide participants with opportunities to engage in real-life settings and reflect upon their outreach skills and help build confidence.
- Ensure access to professional design and production tools and resources, or easy-to-use online resources or softwares.
- Focus students' efforts on real-life products, not classroom exercises.
- Ensure exposure to a wide range of outreach tools and formats to cater to the wide variety of students' needs, abilities, and expectations.

## Conclusion

Overall, our training programme facilitated quantum physics PhD students' acquisition of skills using a pluralistic approach where students attended workshops and created a variety of outreach products over a timeline of nearly two years. This long timeline allowed us to create a rich learning experience and enhanced the quality of students' products through brainstorming and feedback sessions. The programme culminated in the creation of an outreach portfolio, which to our knowledge is first of its kind in outreach training. Furthermore, the training programme provided them with an increased understanding of the importance, benefits, and applicability of outreach skills.

Although the training programme was designed for PhD students in quantum physics, the methodology and findings are domain-agnostic and can be transferred to other contexts and disciplines. Therefore, based on our experience with and assessment of the training programme, this programme can be implemented in universities and research networks to the benefit of early-stage researchers in STEM and society at large.

## Acknowledgements

We acknowledge funding from the European Union's Horizon 2020 research and innovation programme under the Marie Skłodowska-Curie QuSCo grant agreement No. 765267. We thank QuSCo's former Project Officer, Athina Zampara, for her support and encouragement during the programme. We are also grateful to Prof. Christiane Koch, Prof. Steffen Glaser and Prof. Frank Wilhelm-Mauch for contributing to the development of the programme. Furthermore, we are grateful to QuSCo coordinator Mattia Giardini, participating PhD students, and all the senior researchers of the QuSCo network for their support and participation in this programme. We would like to thank our research assistant in the team, Dóra Verasztó, for her assistance in the literature review and coding of responses. Lastly, we are grateful for the support from Carlos Mauricio Diaz-Nissen for the talk on *Conceptual Representations* and Peter Kjærgaard in digital rendering of the infographics.

# Appendix A: Workshop Descriptions

In this appendix, we describe in detail the different workshops conducted over the course of the training programme, in the order in which they were conducted. For each workshop, we outline the overarching purpose, content, and output.

## Scientific Visualization

Telling the story of academic research can be just as important as conducting the research itself. Visualizations like diagrams, cartoons, infographics, and data representations hold vast potential for telling a story succinctly and effectively (Lankow et al., 2012; Shurkin, 2015; Shanks et al., 2017). As with written communication, it is important to know the goals of the visualization (what story do you want to tell and why) as well as the audience (e.g. your colleagues, other scientific researchers in the field, funders, the general public, children).

In our context, the goal of the workshop was to provide students with the tools, motivation, and confidence for creating visualizations to explain their research to high-school students, a non-academic audience, by creating an infographic (and accompanying caption) to communicate their research. Moreover, the infographic was an element that accompanied their blog writing task (as mentioned under *Blog Writing*).

Additionally, we provided support by bringing in a professional artist to enhance the students' infographics with the goal of producing a more polished final product. While the workshop aimed to give students the tools needed to create their own infographics and visualizations, it is often beneficial to work with a professional artist to produce a final product.

The workshop took about 2 hours and included a presentation and sketching session. The presentation covered basic Gestalt principles (Koffka, 1935), styles of graphics, and guiding questions to ask oneself before, during, and after the sketching process. Students were encouraged to 'just keep drawing' during the session and ask the instructor for feedback or to help convey 'characters' (objects, actions, outcomes) in their story. After the session, students were given one week to finish the infographic. The finished infographic sketches and captions were then sent to a professional artist for touch-ups.

The infographics are available on the QuSCo website as part of the students' outreach portfolios. Students were encouraged to use their infographics in other settings and formats such as in presentations, posters and popular science articles, and as explained in the main text, some used them in subsequent presentations.

## Blog Writing

Blogs are considered to be one of the most popular science communication and outreach tools for graduate students and researchers (Brown & Woolston, 2018; Saunders et al., 2017). Moreover, blog writing is one of the most common focuses in various training programmes as skills in written communication can be useful in different scenarios both for science communication and in one's professional career.

The goal of this workshop was to help students write about their research for an audience outside academia. We believe this is important as early-stage researchers and scientists typically write academic articles and communicate their research to their peers. However, communicating with non-experts about their research is required, e.g. when communicating with undergraduate students or potential funders, and students typically face difficulties with it (Baram-Tsabari & Lewenstein, 2012). Moreover, students are also used to writing in a style suitable for academic papers, e.g. using domain-specific terms, which are considered unsuitable for blogs. Therefore, we intended to provide them with strategies for communicating with high-school students. This included training on providing analogies, use of the active voice, developing appropriate headlines, and a focus on the key takeaway message.

The entire duration of the workshop was 2 hours, covering a presentation and a writing exercise. The presentation outlined strategies for writing blogs, drawn heavily from Baram-Tsabari and Lewenstein's work (Baram-Tsabari & Lewenstein, 2012). In the writing exercise, the students outlined a structure and received feedback from the instructor for a blog on the 'What', 'Why' and 'How' of their research, targeted to be readable by high-school students. Additionally, the infographic from the workshop mentioned in the previous section also accompanied the blog. Students were then given two weeks to complete the blog post, followed by the aforementioned two-step feedback process (see main text). During the round of peer feedback, the instructors provided students with feedback pointers to streamline the feedback, e.g. identify words or statements that were difficult to comprehend for a wider audience and needed an analogy or if the take-away message was clear in the blog.

The coordinator of the QuSCo network ran a 'Blog of the Week' campaign, where a different student's blog was shared on Twitter every week. We also encouraged the students to publish their blog on their respective research group websites and add the link in their email signatures.

## Simulations

Visualization and numerical simulation tools in quantum physics are indispensable in modern-day quantum research and education. Therefore, designing simulations for educational outreach settings was an integral part of the QuSCo research network outreach training.

The overarching goal was to allow students to highlight the core of their research through dynamic visualizations and text descriptions using two quantum simulation tools created by two research groups within the QuSCo network: *Quantum Composer* and *SpinDrops*. We introduced *Quantum Composer* as it was designed by the ScienceAtHome team, and this section covers the details of the workshop part using *Quantum Composer*. The introduction and tasks around *SpinDrops* was covered by a different team, but that presentation had the same goals and framework as ours.

The session took about 2 hours and contained two components. First, students were given a light showcase tutorial on using *Quantum Composer,* including a simple slideshow with a brief motivational introduction, followed by small, representative sample tasks relevant to quantum mechanics accompanied by live-coding demonstrations implementing these in the tool. Each task also had one or two related questions discussed briefly among pairs of students and then *in plenum*. The second component was an open-ended hands-on exercise.
As a result of this workshop, students successfully applied the tool to demonstrate a part of their research.

The final product, which became part of their outreach portfolio, was a one-minute video (Figure 14) that they created using a screen-recording software with basic editing features that allowed them to add captions and subtitles describing their simulation. These videos were aimed at undergraduate students and we encouraged the students to use their videos in dissemination and teaching activities.

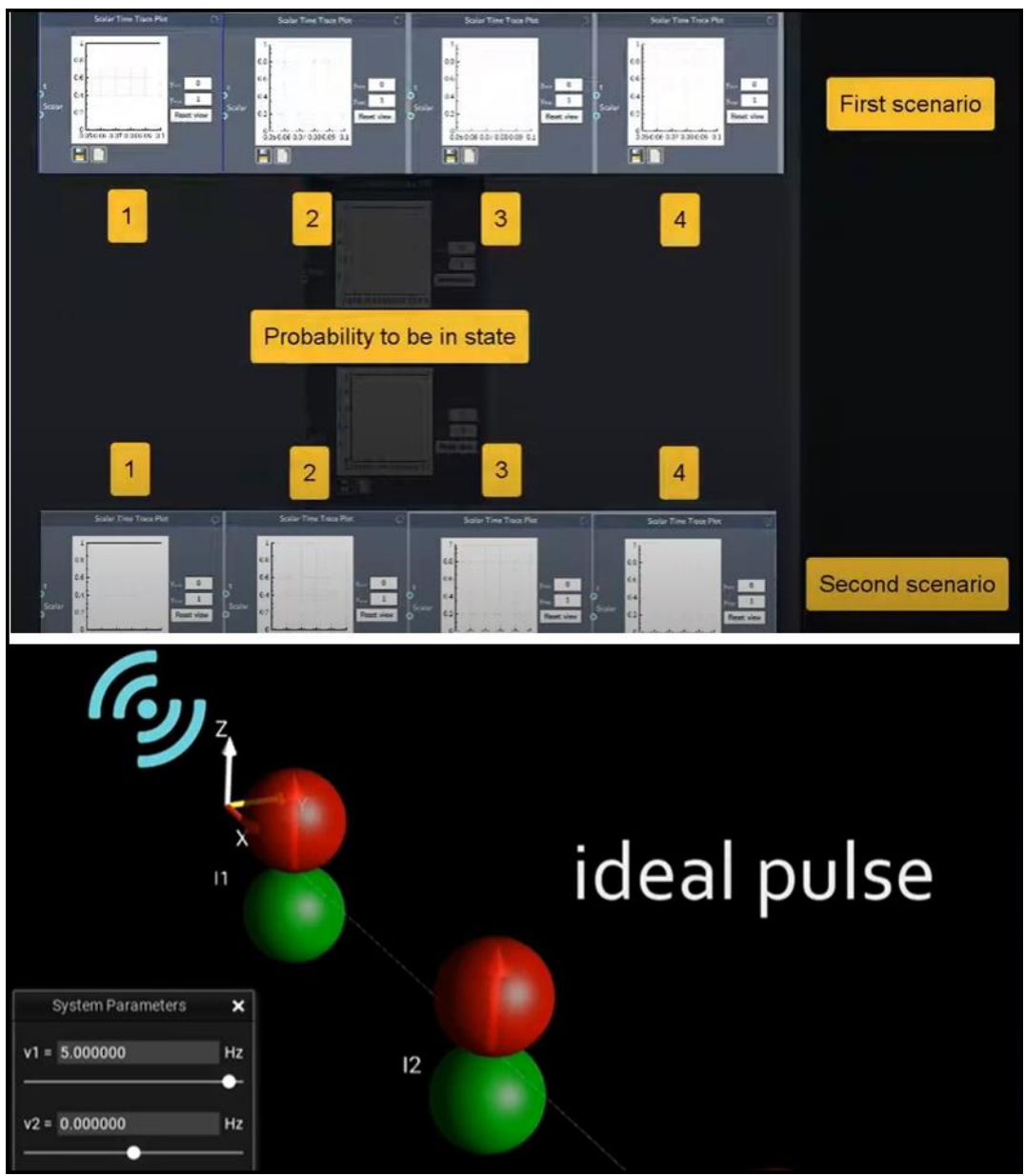

**Figure 14**. Screenshots of Emanuele Albertinale's simulation created in *Quantum Composer* (top) and Marwa Garsi's simulation (bottom) created in *SpinDrops*. These describe an atom's probability to be in different energy levels and how shaping microwave pulses can offer better control of spin systems, respectively.

## Educational Outreach

In order to build on the students' experience with simulation tools, the next workshop focused on the development of educational outreach activities using existing collaborative tools like *SpinDrops* and *Quantum Composer*. The goal of the activity was for students to develop an outline of an educational outreach activity with the assistance of these tools. In this way, students would develop their skills in adapting these tools in ways that can be used to communicate their research.

As part of this activity, we introduced a new tool to the students as added inspiration for the activities they later designed. This tool, known as *SciNote*, is a collaborative problem solving and argumentation tool developed at Aarhus University (Rafner et al., 2021). SciNote, which is built on Webstrates (Klokmose et al., 2015), allows users to explore a problem and publish mini-papers on the problem; these mini-papers can contain data, analyses, or meta-analyses. Other users can then take these data and arguments and build on them in subsequent mini-papers. Thus, SciNote is a useful and interesting outreach tool for teaching scientific inquiry via collaboration within groups of students, e.g. in a classroom setting.

In this workshop, an early version of SciNote was integrated with Quantum Composer. The workshop ran over 2 hours with the goal of introducing the students to SciNote and its capabilities through an example activity around a quantum physics challenge relevant to their work before asking them to design an educational or outreach workshop of their own using SciNote, Quantum Composer, and/or any other tool. In the workshop, students were onboarded with a 30-minute presentation describing SciNote and the quantum physics challenge itself, and they were also provided with PDF documents describing how to log in and use SciNote to publish their data and analyses. Then, for the next hour, either solo or in pairs, the students explored the quantum physics challenge using both Quantum Composer and SciNote.

The workshop finished by asking the students to brainstorm the design of a workshop of their own; a template was provided for them that allowed them to indicate which aspects of SciNote they would use in such an activity and what their proposed activity would cover. There was no obligation placed on the students that they actually implement the activity (and as of the writing of this manuscript, none have done so); the design session allowed them to explore how they would use tools like SciNote and Composer in an outreach activity. Most students designed activities using SciNote, and the target audiences were typically high-school or early undergradaute students. Their proposed activities ranged from solving computer programming challenges using IBM's Quantum Experience to explore quantum computing, to using Composer to understand how to control quantum systems.

## Audio and Visual Communication

Posters, videos and elevator-style pitches are popular and effective means of communicating scientific ideas either visually or orally with any audience to build curiosity or help them retain relevant information in a short period of time (Clarkson et al., 2018; O'Keeffe & Bain, 2018). Inspired by these methods, we conducted a workshop on audio and visual communication where students recorded a video of themselves describing their research or a concept in 15 seconds and created a poster. Designing or creating such outputs are typically non-trivial, as one has to focus on key take-way messaging that can be conveyed in that short period of time. Novices in outreach often struggle with these elements (Brownell et al., 2013). Hence, the goal with our workshop was to train students to convey their main message quickly and effectively in a short period of time using both oral and visual methods. In the process, students also created two dissemination tools: a poster and a video, which can be used in real-world settings, such as science fairs, exhibitions, booths, or in social media.

The 2-hour workshop commenced with the instructors showing examples of posters, graphics and videos from different platforms, e.g. *Physics Reimagined* and *Compound Interest,* to inspire students and provide them with a starting point for their own posters. Prior to poster making, we held a brainstorm session where students discussed what kind of content should be present in their poster, keeping in mind use-case scenarios such as a public outreach event or a lab tour. For example, students discussed ideas on a theme such as showing the working principle of an experimental technique to show during a lab tour or explaining a quantum mechanical concept through analogies at a science fair. Students were encouraged to also think about reusing or adapting some of their own outreach products (such as the blog, video and infographic) so they did not have to start from scratch thinking about the content. Moreover, we also encouraged the students to use online material (with acknowledgement) that was shown during the workshop and add a QR code to send readers to the students' blog or video for further details. At the end of the brainstorming session, we provided the students with a list of guidelines for them to consider during the poster creation task, which was completed after the workshop within two weeks. These guidelines included pointers such as mentioning the novel or challenging aspects of their research and organizing the information in a clear and concise manner similar to the examples shown.

After the poster onboarding, we introduced the video activity where students had to talk about their research or what research meant to them in 15 seconds. The short time requirement was imposed to mirror an elevator-style pitch and prepare students to participate in the *#MyJobinResearch* video challenge on Twitter launched by the European Commission with the goal of informing the public about the work of EU-funded researchers. It was mandatory for students to create a video as part of their portfolio but participation in the challenge was voluntary. Students were divided into groups of three where they first got two minutes to prepare what to say to their peers and then subsequently attempted to talk about their research for only 15 seconds. At the end of the session, students were given tips. For example, they were encouraged to choose one (and only one message) to be conveyed in the video. We also encouraged them to use props to make the video catchy and interesting. Later, students made an audio-visual recording using tools that they were already familiar with (e.g. their phone).

Two weeks after the workshop, students created a ready-to-use poster and video after implementing changes from the feedback they received from peers and experts.

## Inquiry-Based Learning

Understanding science means simultaneously understanding the scientific phenomena being investigated, and the scientific methods with which new knowledge is produced (NGSS Lead States, 2013). Research shows that students who learn scientific core concepts while also learning scientific methods learn both more effectively (NGSS Lead States, 2013). The didactic approach known as *Inquiry-Based Learning* (IBL) focuses on facilitating learners while they engage in scientific inquiry when trying to answer a big, overarching *driving question*, thus combining scientific-inquiry with scientific concepts (Edelson et al., 1999; Loh et al., 2001). As part of our multifaceted approach to preparing participants for engaging the public in science, we therefore wanted to introduce them to inquiry-based learning design.

The aim of the 2-hour IBL workshop was for students to learn about Inquiry-Based Learning by designing a 1-2 hour curriculum unit aimed at high-school or undergraduate students, in which younger students could engage in IBL activities relating to the participant's research. During the workshop, we first took participants through a brief IBL-activity, focusing on the following driving question: how did we get so many differently looking wolves across the world? We deliberately designed the curriculum to be STEM-related, but not related to any research done by our participants, as this would introduce the general IBL-concepts without leading anyone down particular lanes of design of their own work. Participants then spent 45 minutes doing typical IBL-activities: discussing and articulating hypotheses about the driving question; discussing what kind of data would support or reject such a hypothesis; finding some data online to back this up; and finally presenting to each other what they believed would be a good answer to the question, and critiquing each other's presentations. The purpose was for students to not just learn about IBL, but how to *do* IBL.

After this exercise, we presented participants with a design template that they could use to design an IBL-curricular activity of their own. We recommended students to build their activity around the visualization and simulation tools typically used in physics education (e.g. *PhET* (Mckagan et al., 2008)*, QuVis* (Kohnle et al., 2012)*, Quantum Composer, SpinDrops*) as they were already familiar with them from previous workshops. Moreover, simulation tools can provide opportunities for inquiry through manipulation of variables and parameters (Mckagan et al., 2008); we described two concrete examples using *Quantum Composer* in the previous sections. We facilitated an iterative design process for participants: first, they had one week to write up a proposal for their learning activity. The template for the draft included questions like, 'What is the driving question of your activity?', 'What are the learning goals of your activity?', 'What kinds of data will you let students collect, and how do you expect them to analyze and make sense of it?'. Afterwards, we gave them feedback on their draft, and they then submitted a revised, final version two weeks later which was included in their outreach portfolio.

## Outreach Portfolio

Portfolios are collections of relevant products that are used to show and document one's competences and experiences (e.g. in the fields of teaching and design). Typically, portfolios accompany a résumé and are used in job application settings, but they are also used in networking or collaboration settings. Therefore, in order to document and show the students' outreach work, experiences and competences, the training programme culminated with building an outreach portfolio, which can be used by the students in various contexts in the future.

In this 2-hour workshop, the overall goal for the session was to make students aware of the advantages of having an outreach portfolio and to provide a support system to help them build their own. The emphasis was on highlighting the components of an outreach portfolio through some concrete examples and a collective brainstorming activity.

The presentation began by showing a real example of a teaching portfolio of a senior researcher from the ScienceAtHome team (J.S.) that was used during the hiring process for an Associate Professor position. This was followed by another example made by S:Z:A that described the different parts of an outreach portfolio, which consisted of the following parts: (a) Introduction: Who am I?, (b) My motivation: Why is outreach important to me?, (c) My outreach philosophy: My approach and techniques, (d) Outreach products: My collection of products, and (e) Outreach activities: Timeline of conducting or participating in outreach activities. The purpose of showing these examples was to lay the foundation for the students as inspiration for their own portfolio. For the sections on outreach motivation and philosophy, a collective brainstorming activity was conducted where we crowdsourced answers from the students on their outreach philosophies and motivations to do outreach. This also served as an implicit process of reflection on their attitudes and what works for them personally and academically while doing outreach. For example, during the discussion about students' personal motivation to do outreach, some of the responses included desires to "*eliminate scientific misconceptions*" and "*extend my personal CV*". In terms of outreach philosophies, students mentioned that they like to "*share humanistic and personal elements"* and "*use analogies*" while doing outreach. At the end of the session, a number of ready-to-use online resources like *Slidesgo* (https://slidesgo.com/) and recommendations (e.g. templates, visual elements and icons) were also provided in order to help make their portfolios visually appealing.

# Appendix B: Student Survey exported from *SurveyXact*

**Which stage of your PhD are you in?**
- 3
- project stage
- 3rd year
- Last year of PhD
- 2nd year
- 3rd year
- 3rd year
- 2nd year
- Close to final
- 3rd year running
- Manuscript writing
- I'm in the final stage of my Ph. D.

**At this point, what do you think you would like to do after your PhD?**

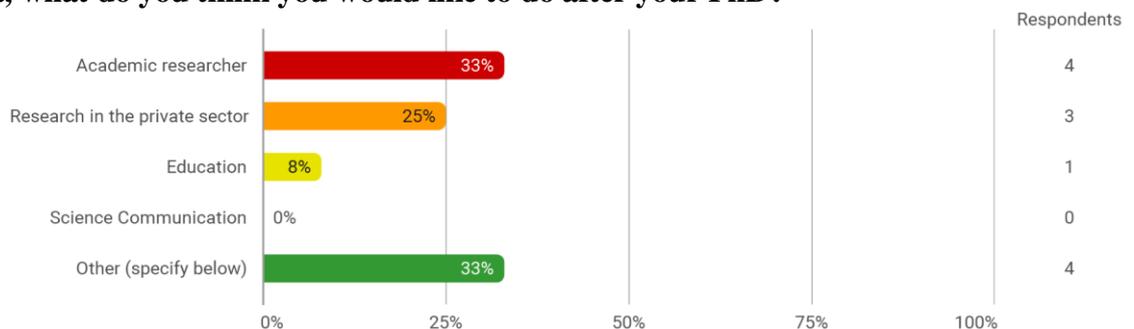

**At this point, what do you think you would like to do after your PhD? - Other (specify below)**
- I am consideirng both the academy and the private sector
- take a break
- Starting my own company
- Possibly some tech companies, not very sure yet

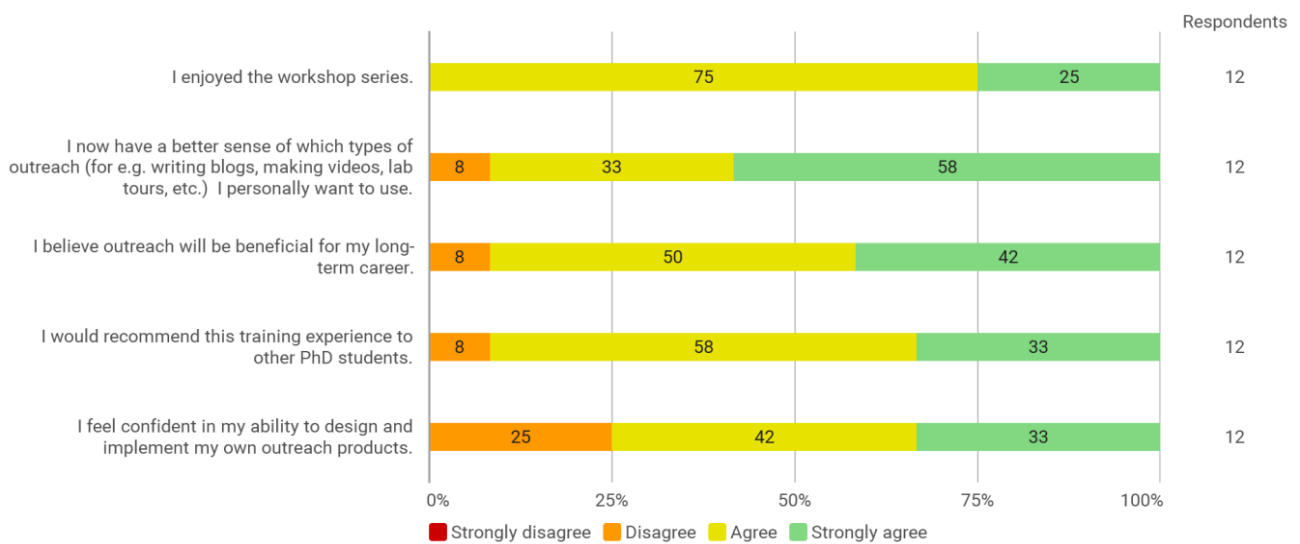

**Each workshop was designed to contain a mix of theory, brainstorming, peer/expert feedback, etc. Select the two that you, in general, found most useful for you.**

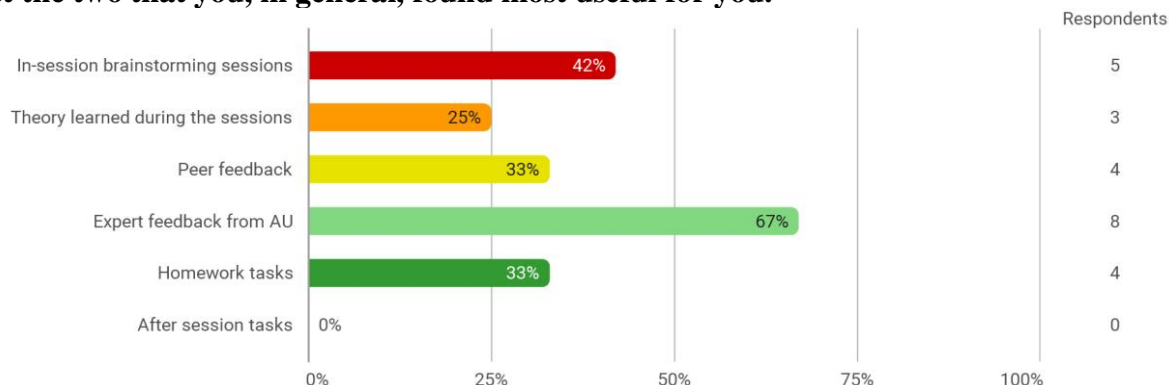

**Tell us about your experience with one of those two elements that you particularly enjoyed.**

- I learnt a lot form the feedback from AU. I found really interstaing see what people thought about my outreach product and how to modify them in order to obtain a better result. The quality of the comment really helped me to understand the mistakes I did.

- The expert feedback helped me to improve my blog post, my poster, and my worksheet. Indeed they spotted some mistakes or rather potential sources for misunderstanding that I had not considered before. I will use the feedback in the future to make sure I do not make the same mistakes again.

- I found the different approaches from my fellow students very interesting, because usually it helped me see perspectives that otherwise I could have not even thought about. So in that regard, that was very useful.

- Brainstorming sessions! I really enjoyed this format because we can just make any proposition we have in mind, hear the others' ones, then get more ideas inspired by their propositions and finally do a fine selection. I believe it's the best way to be creative!
Also, Jacob's feedbacks were really pertinent and helped me a lot in general, e.g. when he gave us feedback the first time we've practiced the 15-sec challenges videos.

- Peer feedback - I think that the community of PhD students that we have in QuSCo is pretty unique. I think we've done a good job looking out for everyone and making everyone feel welcome and included. I say this because this means that people will give honest peer feedback that is constructive and useful. People will provide useful and actionable comments for you to improve, this isn't something that I've experienced in many places.

-
    - I enjoyed the 'in-session brainstorming sessions,' as they provide you an opportunity to convey your thoughts and to get to know thoughts from others.
    - 'Peer feedback' was also one of the elements that I liked, which are indeed quite important. Corrections based on the expert's suggestion was one of the key ingredients for me in these activities.
- I would like to tell you about my experience with homework tasks. At the beginning of my outreach training here in the QuSCo network, as I already said, I thought that outreach is not so important to my objectives in research. This implied that also the requirement of doing outreach homework could be a "waste of time", compared to the time that I daily need to find new results in quantum mechanics.
But after having done the first QuSCo video with quantum composer, and after having listened to

Jesper's suggestions and congratulations (I was impressed by his sentence "I have never thought about using quantum composer in your way, that's just amazing") I suddenly felt like I can do something good also here, right in these homework tasks.

- As I didn't have any previous experience with outreach activities, I found it difficult to design outreach products, such a the blog and the learning activity. Feedback from experts was extremely valuable, both in adjusting my language (in the case of the blog) to be more accessible, and in improving my planned questions, in the learning activities.

- I enjoyed the in-session brainstorming session quite a lot, especially in the remote form. It feels like a very interactive and lively activity at the time and also the create the atmosphere of a team activity for some reason.

- Expert Feedback was really useful in honing my outreach products and reflecting on the outreach methodologies that we used. Specially for the interaction based learning session. It really helped me improve my worksheet and include ideas that made the content better.

- I actually found myself inspired by the feedback of AU experts, as their point of view revealed new perspectives and potentialities of my proposed solution to the assigned homework. Sometimes pointing out aspects I hadn't thought about.

- I enjoyed homework tasks along with peer and expert feedback. Homework tasks took time to accomplish but they also helped me to improve my personal skills and boost the level of my knowledge. Feedback from peers and experts was always inspiring and encouraging.

**Were there any topics or skills covered in the workshops that you would have liked us to spend more time on?**

- The Quantum composer and the Spin drop applications. As an exemple, I did my video using spin drop and so I didn't practice enough with the quantum composer. Propbably I would have liked to practice more  with the quantum composer because I found it a really good tool to teach Physics.

- I quite liked the 15s video challenge. It would have been great to go over some basics like lighting, picking a background, regulating the microphone etc or get access to software that can be used to produce subtitles for hearing impared people or insert graphics. I would have also liked to learn how to use a more professional graphics building tool than powerpoint.

- I do not think "targeting the right audience" has been mentioned more than a handful of times, and it is a vital part of creating an outreach activity. See if we want to motivate, or share knowledge or the right purpose of the activity depends on it. It just felt like we were doing things because we were obligated by QuSCo, and that is just sad.

- I would have liked us to spend more time learning how to write a blog (about the different approaches, styles, and so on). We could have studied an example together to see how it is structured, the kind of vocabulary which is used (I mean if there is a bit of scientific jargon (explained of course) or not). These kinds of things.

- Delivering presentations/doing outreach online - given the shift into the online world that we've experienced this year, and for the foreseeable future, it might be nice to have done something about how to minimise the disruption to taking outreach/presentations online.

- 
  - In general, they were all very well organized workshops, but a little more time on 'in-session brainstorming sessions' would have been another plus. (Having more 'in-session brainstorming sessions' for all the workshops).

- At the beginning of my outreach training period here in the QuSCo network, I didn't know much about different ways of teachings. Then in one of our workshop we covered the topic of inquiry-based learning. This was very interesting and impressive to me, because it is something that I have always been wondering about since high-school. What I would have liked to spend more time on are the different approaches that one has in teaching and learning, besides the inquiry-based one. Are there alternatives? Are there examples of communities which are following these paths in teaching?

- I'm very satisfied with the teachings in the workshop. Maybe focusing on a specific activity, for example the enquiry-based learning, and actually organizing an event were we all made our activity at a 1 day event.

- I think the activities and workshops are very well organised and well distributed time-wise. I wish we had more time revise the infographics actually. I don't feel that the infographics captured the essence of my topic, but of course I was the one to blame starting with.

- I personally felt we could have worked more on the basic skills necessary for the outreach activities that we worked on. Through out it felt more like we are pushing to get the work done, rather than improving our outreach skills. The focus could have been on the skill rather than the final product. Towards the end however, the outreach activities did had more learning element to them.

- I would have liked to spend more time on the graphical part of outreach, like creating infographics and presentations with simple and immediate ideas related to my research work. I've also found interesting the short-video part and would have liked to have something on how be effective in communicating through video.

- I would have liked us to stronger emphasize the role of outreach activities in the beginning and maybe also to spend more time on that. It might sound strange but I saw the whole picture, and the pieces of the puzzle assembled in one big picture just after creating the portfolio.

**Are there specific topics or skills that you would have liked to learn about outreach, that were not covered in the workshops?**

- One of the thing I would have liked to learn is how to organise an outreach event. How to decide the spekers, which kind of activities propose and so on. For exemple, I would have liked to learn about the different possible formats, why it is better to chose one format respect to another one, which are the strong and weak points of the different formats...

- What I would have liked to learn is which difficulties one might run into when interacting with Highschool students and how to deal with them. I would have also appreciated a workshop on how to check whether the level of explanation is appropriate (like https://splasho.com/upgoer5/) or a qusco science slam. I would have also liked to actually practice speaking i.e. make a full length outreach presentation and hold it.

- There was not a single mention to graphic design, not pictures, 3D figures or animations. In world where virtual media is so strong these tools should be focused on and it is a shame that there was no time for that.

- I think some tutorials for making designs would have been really nice (on Inkscape to keep it easy)! This would have been interesting to all of us because we can use it to produce figures for outreach as for research.
Also, an interesting workshop would have been about how to represent data (whether we want to write an article for the general public's journals or scientific ones). I felt like the whole outreach training was focused on targeting high school students (and the general public). I would have enjoyed having it a bit more diverse.

- I think that one aspect that the outreach seminars touched on only a little was the interface with industry, most of the outreach tasks involved talking with kids. Given the number of industry partner QuSCo had, and the idea factory workshops, I think it would have been nice to have some crossover between the two. Explaining science ideas to investors or to members of funding boards, that type of thing. I think you need the outreach skills that we've been developing but you need to do the outreach in a different way.

-
    - I would have liked to learn more about public speaking to improve my communication skills. A dedicated workshop about this would have been another plus point.
    - -------------------------------------------
- At the beginning of my outreach training period here in the QuSCo network, I didn't know much about outreach. I personally would have liked to know more about outreach and its impact on society and science. For example, graphs explaining how outreach is making society more aware of science, or how outreach could be good for your career inside or outside academia. I would have liked to know more about outreach history, names of great experts of this fields, or even great scientists who are known for their commitment in outreach. In other words, it is not clear to me at all the historical root and ways which outreach has followed in our world.

- I think more practical help in initiating an outreach activity would have been also great. Things like, how to contact a school or a venue, what laws to take into consideration, how to advertise the event. Maybe a detailed example or a checklist could have helped.

- I think it would be really cool if we can do a joint outreach product in teams and work on a random topic. But I am not sure whether or not this is useful, but think it would for sure be fun.

- The workshops covered a wide variety of topics, I can think of the following in addition that can be interesting for the participants.

    - It can be useful to learn to work on science vlogs and educational videos.
    - Interviewing skills; specifically in the context of interviewing an expert from the field and then publishing the content in print of audio or video
- No, actually I found that overall the workshops were pretty complete, maybe would have desired more material on the above-mentioned subject, but in general, I am satisfied with the global content and topic coverage.

- I would have liked to learn science popularization techniques or approaches. It was allowed to present well-known physical effects in our homework tasks instead of our research. However, I would like to have better skills of science popularization and present solely my own research.

**Do you have any other comments about the training programme? (Answer will not be shared with the PIs)**

- (Does that imply that the other answers will be shared with the PIs?) I really like QuSCo, however it would have been great to have more in person meetings earlier on.

- The last months have been too much. It was not well distributed over the three years of QuSCo, in my opinion.

- In the beginning, I really felt kind of forced to do outreach and that it was the highest priority in the programme. Which was a bit frustrating because as a PhD researcher, you wish to achieve results before outreaching about them. So even if I like outreach, it felt like a duty.
However, I believe having the first outreach's workshop helped a lot to understand why it is important to build an outreach portfolio. So maybe having this workshop right at the beginning would be better!

- I liked the training program and how it was designed and executed. If it was a bit longer (+6 months or +1 year) then it would have been more beneficial in terms of publications, outreach activities, etc.

- Not really :)

- The main comment I have is that outreach is fun (once one is engaged with it) and very interesting, but it's not easy to carry it out when the PI's are not really interested and engaged themselves in it, but they want you to be fully on research. I personally found myself scratching bits of time out of my daily research work to complete outreach activities instead of being fully engaged with them.

- I'm thankful to the team members who organized all these workshops and were with us all this time. I really appreciate the huge work that has been done by them.

**Which portfolio product are you most proud of?**

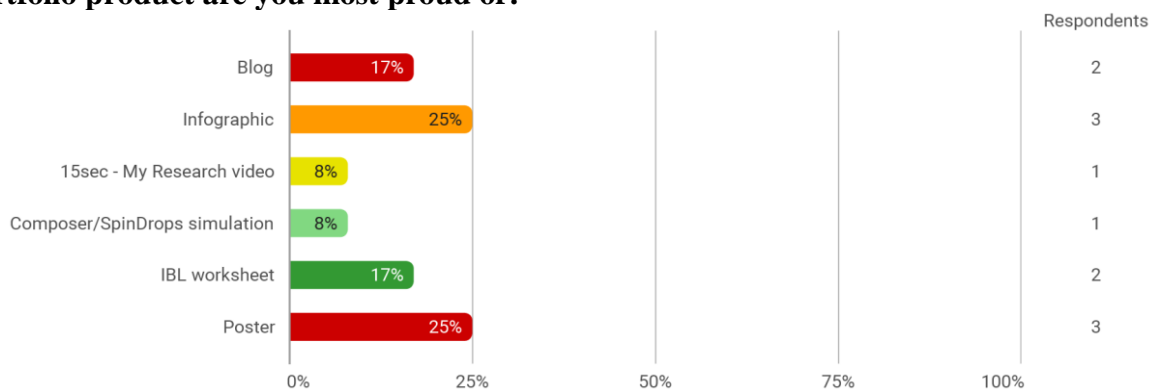

## What about it makes you proud?

- I spent a lot of time doing it and I learnt a lot. Firstly, doing the activity I had the opportunity to better understand some physical concept, secondly I learnt a new way to do a lesson. I think IBL is a powerful and usefeul method that really helps students to better understand difficult concepts.

- I used materials which I had considered in my head for a long time. It was nice to see these things coming together. This is also the only product which people actually get to see (as I have put it on my office door).

- I created it from scratch, even had to create most of the images that are shown there, so all in all I felt like I built it with my own hands (even though I also created the rest of the things similarly). I think I like how it deconstructs ideas and tries to show them in a simple maner.

- Honestly, I really like the render and the analogy I've used (microwave sequences == choreographies ) to build a story! Also, before doing it, I couldn't imagine being able to give insights into optimal control used on my system to people who don't even know about quantum physics. In the end, I realised we don't need people to understand all our research in a video of 1 min, but you need to convey **one** idea!

- I think its fun, accessible and it captures the essence of my research really quickly. It's also absurd and eye catching which makes it a good point to talk with people who are interested in my research.

- For me, the Blog is the outreach product that I am most proud of because it was a challenge for me to explain my research work in simpler terms (targeting bachelor students as an audience), and I think I succeeded in that. And I am pretty sure that this is also going to help me in the future. I also liked the fact that it went through rigorous feedbacks.

- At the beginning of my outreach training programme here in the QuSCo network, I didn't know how to talk about my research. I was not expecting at all that that could be done in just 15 seconds. When I first heard about it, I was of course doubtful. Then I started trying, and I started feeling that this challenge was sort of poetic, because like a poet one should choose only the exact words that you would like to communicate, and each of these words should be a door which opens a world, so that in 15 seconds you can tell about the universe.

- I think I was able to achieve two goals:
  - simplify my topic to be made understandable to the general public
  - make the blog funny and interesting enough to hopefully lead to further questions

- I think it brings out the topic well and actually seems plausible to be used a real worksheet.
  Also I have received good feedback on it in general which I found quite encouraging. It's also fun to make predictions on how students will answer the questions in the worksheet.

- It was an interesting and new activity for me, I liked the sketching session and the final product from the infographic expert team really captured just gist of my imagination for the
  sketch

- It's not only that I'm proud but I enjoyed a lot of thinking and realizing it. Moreover, I think is one of the most creative and immediate methods to reach the public. I found myself forced to think to my research in simpler terms and managed to find creative analogies and illustrations.

- I liked the poster I made since it's written in a simple language understandable to a wide audience and also covers a topic that is not well-known among not scientists. It makes nice connections between the two main topics discussed there.

**Which portfolio product needs most work?**

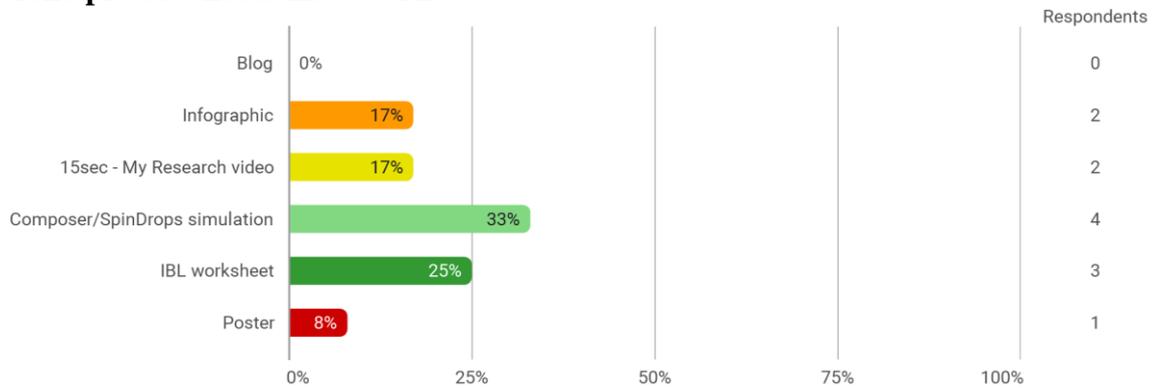

**What about it would you like to change or revise?**

- Meybe i would like to revise the layot of my poster. Probably I would like to add more pictures or introduce some analogies to better expalin my experiment. Moreover it could be nice to add some experimental results and some photos of my set-up

- Well, I realise the video is too fast, there is no background music and the explanation could be better in text and spoken. The way it stands right now, I don't think it is helping people very much.

- I still fail to see how such a short video can be used to show the students anything. Either they know or they don't and the time duration of the video won't be able to change that. I am also not sold on the idea of Composer, but that is just a personal opinion.

- I think I would do the Infographic completely differently today. At that time, I've proposed something because we had to it but I was not 100% convinced. Now I would rather follow the story I have made for the video. Actually, while writing these lines I got the exact idea of what I could draw instead!

- I think I was too ambitious, I had trouble at the time recording the video and then had trouble annotating it properly. The idea behind my research doesn't really come across in 15s as well as I hoped it would. I think now, knowing what I know about outreach, I would do a much better job.

- Explaining your research in a 15-second video is very tricky. I tried my best, but I think I would have done it better.
  --------------------------------------------------------------------------------------------------------

  --------------------------------------------------------------------------------------------------------

- In my inquiry-base learning track I would like my students to understand the origin (and the beauty, why not?) of group theory. As a mathematician, I know very well the struggles of every student who is starting a bachelor in mathematics (or also physics, but in mathematics I did 4 exams of algebra, and an undergraduate physician cannot properly understand this feeling) in facing all these algebraic exams. In mathematics, the origin of group theory was exactly the group of rotations, and this is wonderful because it is physics and engineering. The person who first talked about it was Évariste Galois, a French mathematician from the beginning of the 19th century, who died at the age of 20 in a fight for a woman that he loved. Could you believe it? He discovered one of the most important field of all sciences, and he was only a teenager. Because of that, I would like to add a note about Évariste Galois.

- I think I wasn't able to create interest in the topic, and I underestimated how much you can really convey even in 15 seconds. I was unsure about using a prop, but one well enough designed would have helped

- I don't think it illustrates the core value of my topic well. I would like to make it more eye-catching, by make it more comic like which I only realised after I have seen others. Also I have noticed that camera as the observer is widely used. I would like to avoid this kind of repetition whenever possible, to make each graphics more unique.

- I think IBL was a great learning experience, and it required most of the work because we had to think from a different perspective. I think there could have been few more sessions for this particular task, but otherwise it was great

- I would probably choose a simpler physical phenomenon, more accessible to the wide public instead of a restrained class of people. But it would require a lot more thinking and work (cause simplifying things is never easy)

- I would like to make it even more detailed. There is also need to rephrase the sentences. It may be good to implement the IBL worksheet on a test audience and see how it works. Then it will become clearer how to make it even better.

**Do you have the skills necessary to make these revisions, or do you need more training? If so, what kind of training?**

- I think I have the skills, but it recquires lots of time. For sure, if I had more practice with some desing program it would have been easier. But I think that it is essentialy the lack of time the problem.

- I would need more video editing skills. It would also be nice to talk about ways to make videos easier to watch on the phone and talk to someone from the traget group what they would like to have changed.

- I do have the skills to make these revisions. However in this matter what I lack sometime is a good focus or a good vision of why are we trying to tackle such an activity. It was also a shame that we developed all these tools and yet there was no outreach activity planned by the PIs.

- It depends. If the infographics need to have the same style for all the ESRs, I can't do it. But if I can simply make my own style, it should be fine but it won't render as nicely as if I were a graphist I believe.

- I do have the skills necessary to do it but just dont really have the time, its not something that I plan to show to people, given all the outreach products that we've produced so far.

- I saw some of the videos from other people, and now I have a good idea of doing it better. So I think I don't need any more training specifically related to it.

------------------------------------------------------------------------------------------------------

------------------------------------------------------------------------------------------------------

- Yes I definitely have these skills. I can just add a QR code to his wikipedia page. But I guess that what I would like the most is to talk about his life in a real lesson. I could also suggest a great old Italian movie about him, called "Non ho tempo" (I don't have time). This movie talks about his life and his attempt to communicate his discoveries writing letters to the great mathematicians of that time. One of the most famous of those letters was directed to Cauchy, and it was written right the night before his fight. Since he knew that he was probably going to die, I ended the letter saying that he was sorry about the mess of his notes, but I had no time.

- I think seeing more examples would help, to get a feel of what works. I think I have all the skills now but the most important part would be to invest time, for both thinking about the script and practicing the presenting.

- I think I have the necessary skills for making the revision. By receiving feedbacks and my own reflections, I think I know more about what to emphasise on and how to make the products more appealing.

- Certainly more training/experience with basics outreach methods such as blog, infographic, video and poster. I understand that we did not have enough time to dedicate to these, but hopefully in future I will try to fine-tune my skills further and hopefully come up with better outreach products-

- I maybe need more training on how to simplify concepts and introduce them to the general public. But it would require probably a lot of thinking to find a concept easy and understandable by general public.

- Yes, I think I need more training, I need to work with a real audience and implement my IBL worksheet. In theory it can work but in practice it might go wrong. I hope I will have this opportunity in the nearest future.

**Which part of your portfolio was most difficult to design so that it effectively communicates with non-physicists?**

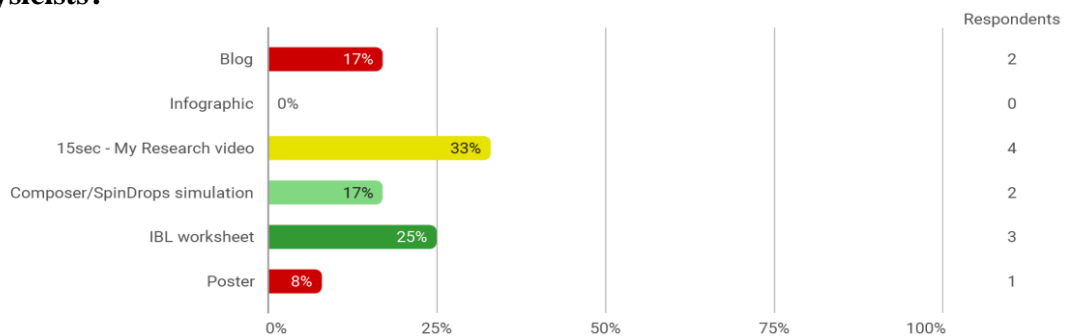

| | Respondents |
|---|---|
| Blog | 17% — 2 |
| Infographic | 0% — 0 |
| 15sec - My Research video | 33% — 4 |
| Composer/SpinDrops simulation | 17% — 2 |
| IBL worksheet | 25% — 3 |
| Poster | 8% — 1 |

## What was most difficult about communicating your research effectively with non-physicists?

- How to introduce the experimental set-up and explain how a two levels system works. I do not know how much it is necessary to say in order to be able to well explain the Physics behind a two levels system whitout being repetitive

- Finding the right tone is tough. I find it simpler to properly curate a short video, a fixed poster or an infographic than a long worksheet which Ihave not tested on anyone yet. In general this type of communication is hard to understand without practice. Even when I present a poster to physicists I need to usually asjust my strategy to make myself clear. With non-physicists that is much harder.

- Explaining very difficult and abstract concepts to people that do not even know what quantum means it is kind of a nightmare. So I felt that in the end what I was going to say was just a very neutral and broad explanation that would differ very little from "I do quantum physics".

- The timing! Before having Jacob's feedback, it was about **explaining** something and **presenting ourselves** in only 15 seconds. Then Jacob explained to us we had to choose one thing to present and we don't have time to explain something.

- In this case I think the big problem was that I was again a bit too ambitious, I wanted to really give people a feel for what actual experimental quantum physics is like, but its tough to remember how little people know about classical physics, let alone manipulating a qubit.

- The blog was the most complicated but the most satisfactory outreach product for me. Since quantum mechanics itself is quite strange and hard to understand, so it was difficult to explain some scientific phenomenon (basic to me) to a non-physicist. But I tried to explain it by providing some hyperlinks to other simplified blogs and videos.

- What is the biggest difficulty for me is that I am not even a physicists. Being a mathematician, everything is even more abstract and hard to explain to non-mathematicians. I do not have real experiments to talk about, my research is all about abstraction. Nevertheless, abstraction is beautiful and peaceful, so I tried very hard to communicate those feelings.

- The poster doesn't allow for a the generation of a slow understanding.
  You have to explain and generate interest in the same slide and this is extremely hard.
  I think one have to accept that it needs to be more catchy, leading to further questions, instead of trying to explain.

- It's hard to foresee what non-physicists consider important and fun to learn about. Also 15second is really short. But most importantly I think I fail to see the core value of my topic in terms of outreach.

- It is really difficult to explain anything considerable to a non-expert in 15 seconds, even by using props or any other
  medium

- I personally do not think that all topics are accessible to the wide public through inquiry-based learning. Many topics require previous knowledge (sometimes advanced) of the physics behind a phenomenon (like quantum physics)

- I have lack of science popularization skills. I have read popular science books. Sometimes, you want to say wow when reading those books which is not the case with my blog. I'd like to make it more fun to read.

**I feel well-equipped to conduct outreach using my portfolio products for the European Researchers' Meet on January 12th, 2021?**

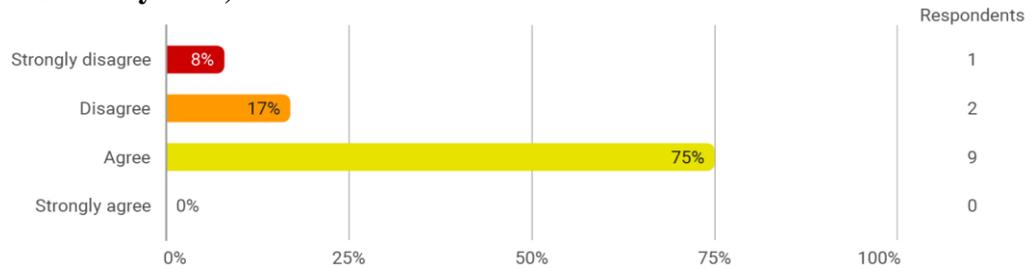

**What are you most excited about?**

- The fact that I have to talk about something different form my experimental results. The possibility to talk about my point of voiew on Outreach. I think that the Phylosophy of the Outreach can be a really interesting starting point for some discussions or some personal reflections.

- Actually testing some of my outreach products will surely be fun. One of my major motivations for outreach is to tell people what there money is spent on as we are funded through the public. SO this is a great opportunity. I have now spent 20min on this survey and need to up my character count. It might be ncie to implement an overall count or something along these lines.

- The portfolio products are very specific tasks and, if I have understood it correctly, will not show the students the things they are looking for! They want to see how my research life is and my products show very simple examples in physics. They have nothing to do with each other! Because the audience is very different! That is what I was talking about before regarding the audience.

- Getting people who know nothing of your research excited about what you are doing by listening to you! It already happened to me during lab tours and we've even got students joining the team. I still feel proud of it! ;)

- Showing off my infographic and hopefully engaging in some interesting discussions about quantum physics. I would like to encourage people to look further into quantum research as a possible career choice.

- I am excited that this would be the first time I will present my research to a large and broader audience. I will also get a chance to interact with the students and discuss perspectives from both sides about research.

- I am most excited about questions and remarks from the audience. I hope they will like my research and my daily life in academia. Questions are always challenging because they are totally unknown. And the feeling of unknown has always been exciting to me, maybe because I am a sort of romantic (in the humanistic sense, like Manzoni or Baudelaire). Moreover, questions are sources of time to talk

- more and most importantly they are sources of new discoveries.

- I'm most excited creating a game that can be played by everyone and it can be used to solve real problems and teach new science to the user.
  I'm also very interested in trying out the inquiry-based learning activity.

- I am quite excited about explain to them what kind of researches are out there in quantum physics. And share with them my experiences as a phd student. Also to get some questions from some actual audience to evaluate how well the presentation is received and to acquire some feedback for improvement for the future.

- Listening to my QuSCo pals talks about their research work in 5 mins.

- I am excited about communicating with younger students to make them passionate about physics and also answer their questions and doubts about the physicist career. (and also use my outreach portfolio)

- I'm really excited to revise my outreach portfolio. I think it can be a benefit when applying for a post-doc. Outreach will also be helpful for me if I continue my career in academics since it forces us to think of a better imagery when speaking about science and this is what it makes more beautiful.

**I think I could use my portfolio in job interviews or for communicating with other physicists**

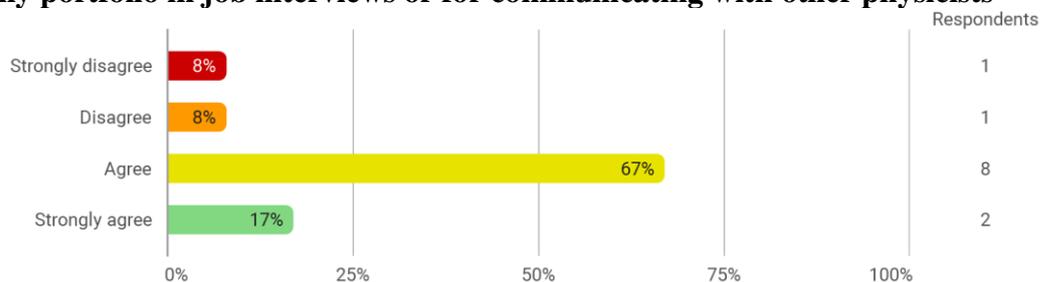

**I believe that outreach skills will help researchers perform better academically.**

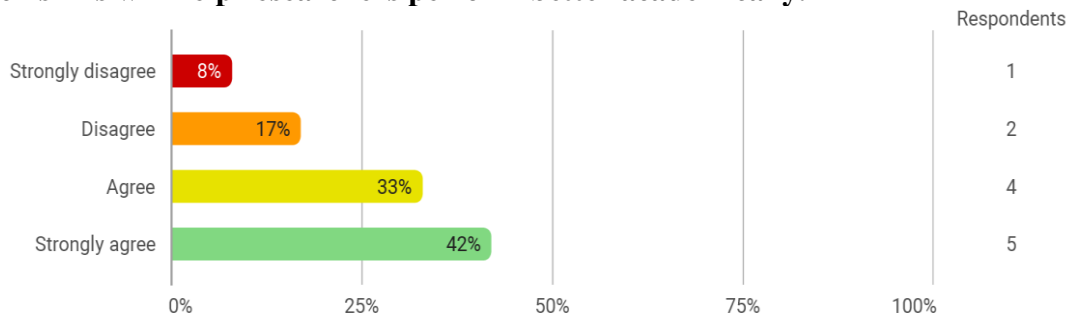

**What were the things you learned about in the workshop series that you least expected to learn about?**

- I liked very much the inquired-based learning seminar and the work we have done to create our inquired-based worksheet. I didn't expect something like this and I really appreciated it. Differently from the blog or the videos, that I liked whitout, however, being motived in doing it, I found very useful and intersting doing the inquired based learning wrksheet. And I think it is an outreach activity that I will use in the future.

- This question is difficult to answer as it presumes that I firstly had specific expectations and secondly remember what they were 2.5 years ago. However, I can say that I learned a number of things I find very useful. For example, I now know how to properly structure blog posts, I have been given certain design tools I wasn't aware of before and I was introduced to IBL. I have built some new metphors and got the opportunity to test them. Furthermore, I am now able to articulate my feelings about outreach (i.e. philosophy and motivation) in a more mainstreamed manner. I believe this will be helpful during future applications procedures.

- After participating in this learning process, I have been able to obtain a lot of communication tools that will help me in my future in a lot of different fields. I have already used them in teaching and in my scientific career, when doing talks in seminars.

- In summary, I would say the fact it can make the difference on a CV. Before having the workshop series, I did not realise the impact outreach can have on our career and how important it is in general. For me, it was something I did in the past because I was simply enjoying sharing and because I wanted to motivate people to join science but without really thinking about it further. During that time I also learned how much time it can consume and I did not really plan to continue for my PhD. But now, with the experience I've got within QuSCo, I am convinced about continuing with outreach, one just needs to find the right timing and of course the right balance.

- I think I was surprised to learn about inquiry based learning activities, I had no idea that this was the name for this type of activity. Throughout my undergraduate degree we regularly did this kind of activity as part of our quantum mechanics lectures and I enjoyed them. Whenever we were asked to think about the type of outreach we wanted to do these lectures came to mind (because I think everyone got good grades in these classes too) but I wasn't sure how to look up information on them, now I am.

- In my personal experience, these outreach activities were quite useful for me. They were properly designed and conveyed appropriately. I learned how to convey my research to people ranging from non-experts to experts. I personally like reading science blogs and getting to know what's new happening in science and these activities helped me to do the same.

- At the beginning of my outreach training here in the QuSCo network, I personally was not feeling confident in talking about my research fields, topics and interests. Moreover, I was even less confident in talking about those issues to a non-scientific audience, to non-experts. By now, after these workshop series, I feel like I could explain quantum mechanics to my grandmother.

  At the beginning of my outreach training here in the QuSCo network, I thought that outreach was not very important for my career in academia or outside academia. I thought that research is all about theorems and experiments, and that there would be no time for anything alse.  By now, after these workshop series, I have realised that the ability of speaking about your research, advertise your scientific results, which also means conveying excitement in science to other people, is as much

- important as your results.

- In the workshop I learned about teaching techniques that can apply to any subject, not just outreach in Science. In particular, I didn't expect to learn how to create an inquiry-based learning activities.

- First thing I learned and without thinking much over it, is how much effort it takes to prepare the outreach products. And during the process, one also reflects lots over the topic one-self.

- The following:

  - Gamification of quantum mechanics based concepts; it was interesting to learn and process of game development and it's application to the research
  - Interaction based learning; This gave me a new perspective to explain something to a general audience via getting them to interact

- I didn't know about inquiry-based learning and found it a very interesting technique, even if probably easier to apply to some subjects than others. I also found it engaging and stimulating to think about science in simpler terms, such that even non-specialists can understand it.

- Before attending the workshop series I had no idea that there were so many job opportunities for scientific workers in industrial companies. I also enjoyed the outreach workshop series. It gave me a global and detailed image of how the outreach portfolio should look like.

## What are you most nervous about?

- The fact that i am not too much confident with the language and I have to speak in public. But this is a problem that I usually have, and I think that the fact that we have to do a video conference, does not help.

- I am unsure of the exact level I should be presenting at. I also need to find someone to practice on and give me feedback. As I do not have these things I am unsure that the presentation will indeed go well as I am sure you can imagine.

- I am nervous about the fact that I have to create a new product for a five minute activity that it will rob me from a lot of my time, and that is the truth with outreach, it usually takes a lot of time that prevents you from doing research at that time, so it is hard to balance.

- Haha, I am nervous about all the amount of work I still need to do aside (I am in the last year of my PhD remember). Regarding the event itself, I am not yet nervous but maybe if no one cares? It would be a pity!
  Also, I would have liked to know a bit more about the event.

- Not coming across as well as other ESRs who have made more engaging outreach portfolio pieces. Alienating people and making them think that my research is actually really boring and putting them off physics.

- I am nervous about the time. Because it will be quite complicated to present everything within 5 minutes.

--------------------------------------------------------------------------------------------------------------------------

--------------------------------------------------------------------------------------------------------------------------

- I am nervous about the timing. I think that five minutes are good and enough, but still when you have to discuss about something complicated in an easy way, it takes more time. So I will probably tried more than a couple of times, just to be sure.

- I'm very nervous about the questions that I will be asked, very simple questions from people outside of physics tend to make me really question my own understanding.
  Though this is also exciting because I think this is how you get better at explaining and also get a deeper knowledge.

- I am most nervous about there will be questions that I don't know how to answer or that I fail to portrait research as a fun thing to do :). I want them to see the best of research environment and so on, but at the same time, as we all know, it's not always easy in research. So I am feeling a bit unsure on exactly what to share.

- Nothing in
  particular

- I am most nervous about not being able to efficiently communicate the ideas behind my research work or being too specific and technical and not engage the students. I hope being able to empathize with them.

- I'm really nervous about the European researchers' meeting. Emotionally it's hard for me to speak to an audience especially when it's online. I hope there won't be any technical problems and the meeting will pass smoothly.

# Appendix C: Senior Researcher Survey exported from SurveyXact

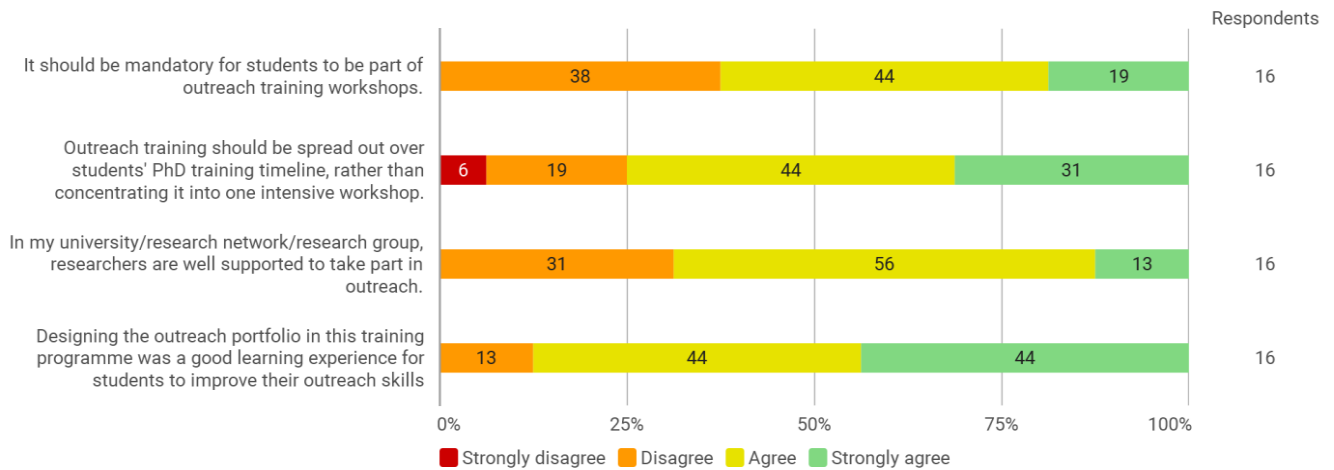

**Have you or could you imagine encouraging your PhD student to use products from their portfolio for non-outreach settings as well? If yes, which setting(s)?**

- yes, for instance to explain their scientific results to other physicists who do not work on quantum technologies.

- participation in science slams
  participation in open lab days
  university ambassador to high schools

- I haven't explicitly encouraged them, but I guess their products could be useful e.g. for job interviews, if that's what you mean.

- Yes. Open houses

- -

- Planning to use some material for our homepage

- I haven't, but I think it would make sense

- -

- Yes, possibly conferences.

- Yes, in introductory part of seminars for instance.

- In addition to outreach to the general public, to schools and the general scientific community, the products of their portfolios or elements thereof will also be useful in scientific papers, conference talks and conference posters for the more specialized community as well as in funding proposals etc.

- The training in outreach could be useful for the students to better value his achievements when he applies for a job or a position.

- Yes, pro-bono initiatives

**Do you have any other comments or suggestions?**

- I think it is important that the students are presented, at the very beginning of the outreach training, with an overview of the what awaits them
  maybe testimonials from students initially reluctant to this training may be helpful in future training programs

- -

- No.

- The outreach training program within QuSCo was extremely well designed and implemented and the success of this program was impressively demonstrated in the recent presentations of the individual outreach portfolios given by the doctoral students.

# References

Andrade Oliveira, L. M., Bonatelli, M. L., & Abreu Pinto, T. C. (2019). DivulgaMicro: A Brazilian Initiative To Empower Early-Career Scientists with Science Communication Skills. *Journal of Microbiology & Biology Education*, *20*(1). https://doi.org/10.1128/jmbe.v20i1.1616

Baram-Tsabari, A., & Lewenstein, B. V. (2013). An Instrument for Assessing Scientists' Written Skills in Public Communication of Science. *Science Communication, 35*(1), 56–85. https://doi.org/10.1177/1075547012440634

Basken, P. (2009). Often distant from policymaking, scientists try to find a public voice. *Chronicle of Higher Education* 55: 38. https://www.chronicle.com/article/often-distant-from-policy-making-scientists-try-to-find-a-public-voice/

Bass, E. (2016). The Importance of Bringing Science and Medicine to Lay Audiences. *Circulation 133*(23), 2334–2337. https://doi.org/10.1161/CIRCULATIONAHA.116.023297.

Bishop, L. M., Tillman, A. S., Geiger, F. M., Haynes, C. L., Klaper, R. D., Murphy, C. J., Orr, G., Pedersen, J. A., DeStefano, L., & Hamers, R. J. (2014). Enhancing Graduate Student Communication to General Audiences through Blogging about Nanotechnology and Sustainability. *Journal of Chemical Education*, *91*(10), 1600–1605.

Bobroff, J., & Bouquet, F. (2016). A project-based course about outreach in a physics curriculum. *European Journal of Physics*, 37(4), 045704. https://doi.org/10.1088/0143-0807/37/4/045704

Brown, E., & Woolston, C. (2018). Why science blogging still matters. *Nature*, 554(7690), 135–137. https://doi.org/10.1038/d41586-018-01414-6

Brownell, S. E., Price, J. V., & Steinman, L. (2013). Science Communication to the General Public: Why We Need to Teach Undergraduate and Graduate Students this Skill as Part of Their Formal Scientific Training. *Journal of undergraduate neuroscience education: JUNE: a publication of FUN, Faculty for Undergraduate Neuroscience*, *12*(1), E6–E10.

Clark, G., Russell, J., Enyeart, P., Gracia, B., Wessel, A., Jarmoskaite, I., Polioudakis, D., Stuart, Y., Gonzalez, T., MacKrell, A., Rodenbusch, S., Stovall, G. M., Beckham, J. T., Montgomery, M., Tasneem, T., Jones, J., Simmons, S., & Roux, S. (2016). Science Educational Outreach Programs That Benefit Students and Scientists. *PLoS Biology*, *14*(2). Scopus. https://doi.org/10.1371/journal.pbio.1002368

Clarkson, M. D., Footen, N. R., Gambs, M. F., Gonzalez, I. F., Houghton, J., & Smith, M. L. (2014). Extended Abstract: Engage, A Model of Student-Led Graduate Training in Communication for STEM Disciplines. *2014 IEEE International Professional Communication Conference (IPCC)*, 1–2. https://doi.org/10.1109/IPCC.2014.711114

Clarkson, M. D., Houghton, J., Chen, W., & Rohde, J. (2018). Speaking about science: A student-led training program improves graduate students' skills in public communication. *Journal of Science Communication*, *17*(2). https://doi.org/10.22323/2.17020205

Cohen, J. (1960). A Coefficient of Agreement for Nominal Scales. *Educational and Psychological Measurement*, *20*(1), 37–46. https://doi.org/10.1177/001316446002000104

CORDIS. (2020). *Quantum-enhanced Sensing via Quantum Control.* European Comission. *https://cordis.europa.eu/project/id/765267*

Crone, W. C., Dunwoody, S. L., Rediske, R. K., Ackerman, S. A., Petersen, G. M. Z., & Yaros, R. A. (2011). Informal science education: A practicum for graduate students. *Innovative Higher Education*, *36*(5), 291–304. https://doi.org/10.1007/s10755-011-9176-x

Davies, S. R. (2013). Research staff and public engagement: A UK study. *Higher Education*, *66*, 725–739. https://doi.org/10.1007/s10734-013-9631-y

Edelson, D. C., Gordin, D. N., & Pea, R. D. (1999). Addressing the challenges of inquiry-based learning through technology and curriculum design. *Journal of the learning sciences*, *8*(3-4), 391-450.

European Commission. (n.a.). *Outreach and Communication Activities in the MSCA under Horizon 2020.* https://ec.europa.eu/assets/eac/msca/documents/documentation/publications/outreach_activities_en.pdf

Finan, E., Mathis, H., Hennig, A., & Nofziger, M. (2018). Development of an online hub for OSC outreach efforts. In G. G. Gregory (Ed.), *Optics Education and Outreach V* (Vol. 10741, p. UNSP 107410Y). Spie-Int Soc Optical Engineering. https://doi.org/10.1117/12.2322089

Fogg-Rogers, L. A., Weitkamp, E., & Wilkinson, C. (2015). *Evaluation of the Royal Society Education Outreach Training Course*. https://uwe-repository.worktribe.com/output/834100

Gau, K. H., Dillon, P., Donaldson, T., Wahl, S. E., & Iwema, C. L. (2020). Partnering with postdocs: A library model for supporting postdoctoral researchers and educating the academic research community. *Journal of the Medical Library Association*, *108*(3), 480–486. https://doi.org/10.5195/jmla.2020.902

Gianaros, P. J. (2006). A seminar on scientific writing for students, postdoctoral trainees, and junior faculty. *Teaching of Psychology*, *33*(2), 120–123.

Heath, K. D., Bagley, E., Berkey, A. J. M., Birlenbach, D. M., Carr-Markell, M. K., Crawford, J. W., Duennes, M. A., Han, J. O., Haus, M. J., Hellert, S. M., Holmes, C. J., Mommer, B. C., Ossler, J., Peery, R., Powers, L., Scholes, D. R., Silliman, C. A., Stein, L. R., & Wesseln, C. J. (2014). Amplify the Signal: Graduate Training in Broader Impacts of Scientific Research. *Bioscience*, *64*(6), 517–523. https://doi.org/10.1093/biosci/biu051

Holliman, R., & Warren, C. J. (2017). Supporting future scholars of engaged research. *Research for All*, *1*(1), 168–184. https://doi.org/10.18546/RFA.01.1.14

Klokmose, C. N., Eagan, J. R., Baader, S., Mackay, W., & Beaudouin-Lafon, M. (2015, November). Webstrates: shareable dynamic media. In *Proceedings of the 28th Annual ACM Symposium on User Interface Software & Technology*, 280-290. https://doi.org/10.1145/2807442.2807446

Koffka, K. (1935). *Principles of Gestalt Psychology*. New York: Harcourt, Brace and Co.

Kohnle, A., Bozhinova, I., Browne, D., Everitt, M., Fomins, A., Kok, P., Kulaitis, G., Prokopas, M., Raine, D., & Swinbank, E. (2013). A new introductory quantum mechanics curriculum. *European Journal of Physics, 35*(1), 015001. https://doi.org/10.1088/0143-0807/35/1/015001

Kuehne, L.M., Twardochleb, L.A., Fritschie, K.J., Mims, M.C., Lawrence, D.J., Gibson, P.P., Stewart-Koster, B. and Olden, J.D. (2014), Practical Science Communication Strategies for Graduate Students. Conservation Biology, 28: 1225-1235. https://doi.org/10.1111/cobi.12305

Lankow, J., Ritchie, J., & Crooks, R. (2012). *Infographics: The power of visual storytelling*. John Wiley & Sons.

Loh, B., Reiser, B. J., Radinsky, J., Edelson, D. C., Gomez, L. M., & Marshall, S. (2001). Developing reflective inquiry practices: A case study of software, the teacher, and students. *Designing for science: Implications from everyday, classroom, and professional settings*, 279-323.

Marbach-Ad, G., & Marr, J. (2018). Enhancing Graduate Students' Ability to Conduct and Communicate Research through an Interdisciplinary Lens. *Journal of Microbiology & Biology Education*, *19*(3). https://doi.org/10.1128/jmbe.v19i3.1592

McCartney, M., Childers, C., Baiduc, R. R., & Barnicle, K. (2018). Annotated Primary Literature: A Professional Development opportunity in Science Communication for Graduate Students and Postdocs. *Journal of Microbiology & Biology Education*, *19*(1). https://doi.org/10.1128/jmbe.v19i1.1439

McKagan, S., Perkins, K., Dubson, M., Malley, C., Reid, S., LeMaster, R., & Wieman, C. (2008). Developing and researching PhET simulations for teaching quantum mechanics. *American Journal Of Physics*, *76*(4), 406-417. https://doi.org/10.1119/1.2885199

Mellor, F. (2013). Twenty years of teaching science communication: A case study of Imperial College's Master's programme. *Public Understanding of Science, 22*(8), 916–926. https://doi.org/10.1177/0963662513489386

Miller, S., & Fahy, D. (2009). Can Science Communication Workshops Train Scientists for Reflexive Public Engagement? The ESConet Experience. *Science Communication*, *31*(1), 116–126. https://doi.org/10.1177/1075547009339048

NGSS Lead States. National Research Council. (2013). *Next Generation Science Standards: For States, By States*. Washington, DC: The National Academies Press. https://doi.org/10.17226/18290.

National Science Foundation. (2014), *Perspectives on Broader Impacts.* *https://www.nsf.gov/od/oia/publications/Broader_Impacts.pdf*

Neeley, L., Goldman, E., Smith, B., Baron, N., Sunu, S. (2014, January 1). GradSciComm Report and Recommendations: Mapping the Pathways to Integrate Science Communication Training into STEM Graduate Education. https://www.informalscience.org/gradscicomm-report-and-recommendations-mapping-pathways-integrate-science-communication-training

Noble, D. B., Mochrie, S. G. J., O'Hern, C. S., Pollard, T. D., & Regan, L. (2016). Promoting Convergence: The Integrated Graduate Program in Physical and Engineering Biology at Yale University, a New Model for Graduate Education. *Biochemistry and Molecular Biology Education*, *44*(6), 537–549. https://doi.org/10.1002/bmb.20977

Nymann, K. (2020, September 11). Prize winning quantum outreach. *ScienceAtHome.* https://www.scienceathome.org/community/blog/prize-winning-quantum-outreach/

O'Keeffe, K., & Bain, R. (2018). ComSciCon-Triangle: Regional Science Communication Training for Graduate Students. *Journal of Microbiology & Biology Education*, *19*(1). https://doi.org/10.1128/jmbe.v19i1.1420

Ponzio, N. M., Alder, J., Nucci, M., Dannenfelser, D., Hiltons, H., Linardopoulos, N., & Lutz, C. (2018). Learning Science Communication Skills Using Improvisation, Video Recordings, and Practice, Practice, Practice. *Journal of Microbiology & Biology Education*, *19*(1). https://doi.org/10.1128/jmbe.v19i1.1433

Rafner, J., Hjorth, A., Weidner, C., Ahmed, S.Z., Poulsen, C., Klokmose, C., and Sherson, J. (2021). SciNote: Collaborative Problem Solving and Argumentation Tool. *Accepted in CSCL 2021*.

Research Executive Agency. (2020, June 9). My Job in Research in 15 seconds. European Commission. https://ec.europa.eu/info/news/my-job-research-15-seconds-2020-jun-09_en

Rodgers, S., Wang, Z., Maras, M. A., Burgoyne, S., Balakrishnan, B., Stemmle, J., & Schultz, J. C. (2018). Decoding Science: Development and Evaluation of a Science Communication Training Program Using a Triangulated Framework. *Science Communication*, *40*(1), 3–32. https://doi.org/10.1177/1075547017747285

Royal Society (2006, June 1). *Science Communication.* https://royalsociety.org/topics-policy/publications/2006/science-communication/

Saunders, M. E., Duffy, M. A., Heard, S. B., Kosmala, M., Leather, S. R., McGlynn, T. P., Ollerton, J., & Parachnowitsch, A. L. (2017). Bringing ecology blogging into the scientific fold: measuring reach and impact of science community blogs. *Royal Society Open Science*, *4*(10), 170957. https://doi.org/10.1098/rsos.170957

Scalice, D., Dolci, W., Brochu, L., Merriman, T., Davis, H., Billings, L., & Voytek, M. A. (2019). FameLab USA: Improving Science Communication Skills for Early Career Scientists. *Astrobiology*, *19*(4), 614–623. https://doi.org/10.1089/ast.2017.1809

Shanks, J. D., Izumi, B., Sun, C., Martin, A., & Byker Shanks, C. (2017). Teaching Undergraduate Students to Visualize and Communicate Public Health Data with Infographics. *Frontiers in Public Health*, *5*, 315. https://doi.org/10.3389/fpubh.2017.00315

Shurkin, J. (2015). Science and Culture: Cartoons to better communicate science. *Proceedings of the National Academy of Sciences*, *112*(38), 11741. https://doi.org/10.1073/pnas.1515144112

Silva, J., & Bultitude, K. (2009). Best practice in communications training for public engagement with science, technology, engineering and mathematics. *Journal of Science Communication, 08*(02), A03. https://doi.org/10.22323/2.08020203

St Angelo, S. K. (2018). Encouraging the Art of Communicating Science to Nonexperts with Don't Be Such a Scientist. *Journal of Chemical Education*, *95*(5), 804–809. https://doi.org/10.1021/acs.jchemed.7b00963

Strang, C., Dorph, R., & Halversen, C. (2005). Communicating Ocean Sciences: A course that improves education public outreach. Proceedings of OCEANS 2005 MTS/IEEE, 3, 2642–2646. https://doi.org/10.1109/OCEANS.2005.1640171

Trautmann, N. M., & Krasny, M. E. (2006). Integrating teaching and research: A new model for graduate education? *Bioscience*, *56*(2), 159–165. https://doi.org/10.1641/00063568(2006)056[0159:ITARAN]2.0.CO;2

Triezenberg, H. A., Doberneck, D., Campa, H., & Taylor, W. W. (2020). Mid- and High-Engagement Programs to Develop Future Fisheries Management Professionals' Skills. *Fisheries*, *45*(10), 544–553. https://doi.org/10.1002/fsh.10480

Vitae. (2011). *Researcher Development Framework.* https://www.vitae.ac.uk/vitae-publications/rdf-related/researcher-development-framework-rdf-vitae.pdf/view

Wang, J. T. H., Power, C. J., Kahler, C. M., Lyras, D., Young, P. R., Iredell, J., & Robins-Browne, R. (2018). Communication Ambassadors-an Australian Social Media Initiative to Develop Communication Skills in Early Career Scientists. *Journal of Microbiology & Biology Education*, *19*(1). https://doi.org/10.1128/jmbe.v19i1.1428

Warrens, M. J. (2015). Five Ways to Look at Cohen's Kappa. *Journal of Psychology & Psychotherapy*, *05*(04). https://doi.org/10.4172/2161-0487.1000197

Webb, A. B., Fetsch, C. R., Israel, E., Roman, C. M., Encarnación, C. H., Zacks, J. M., Thoroughman, K. A., & Herzog, E. D. (2012). Training scientists in a science center improves science communication to the public. *Advances in Physiology Education*, *36*(1), 72–76. https://doi.org/10.1152/advan.00088.2010

Weidner, C.A., Ahmed, S.Z., Jensen, J.H.M., Sherson, J.F., and Lewandowski, H.J. (2020). Investigating student use of a flexible tool for simulating and visualizing quantum mechanics. In *Proceedings of the 2020 Physics Education Research Conference*, 563-568. https://doi.org/10.1119/perc.2020.pr.Weidner

Wortman-Wunder, E., & Wefes, I. (2020). Scientific Writing Workshop Improves Confidence in Critical Writing Skills among Trainees in the Biomedical Sciences. *Journal of Microbiology & Biology Education*, *21*(1). https://doi.org/10.1128/jmbe.v21i1.1843

Zaman Ahmed, S., Jensen, J. H. M., Weidner, C. A., Sørensen, J. J., Mudrich, M., & Sherson, J. F. (2021). Quantum composer: A programmable quantum visualization and simulation tool for education and research. *American Journal of Physics*, *89*(3), 307–316. https://doi.org/10.1119/10.0003396